\documentclass{article}
\usepackage{graphicx}  
\usepackage{amsmath}   
\usepackage{amssymb}   
\usepackage{bm} 
\usepackage{dcolumn}
\usepackage{color}
\usepackage{mathrsfs}
\usepackage{amsfonts}
\usepackage{varioref}
\usepackage{slashed}
\RequirePackage[colorlinks,citecolor=blue,urlcolor=magenta,linkcolor=blue]{hyperref}
\def\be{\begin{equation}}
\def\ee{\end{equation}}

\labelformat{section}{Section #1} 
\labelformat{subsection}{Section #1} 
\labelformat{subsubsection}{Section #1}
\labelformat{subsubsubsection}{Section #1}
\labelformat{equation}{Eq.~(#1)} 
\labelformat{figure}{Fig.~#1} 
\labelformat{subfigure}{Fig.~\thefigure#1} 
\labelformat{table}{Tab.~#1} 
\labelformat{appendix}{Appendix #1}

\title{\bf Fermionic Bell violation in the presence of background electromagnetic fields in the cosmological de Sitter spacetime}
\author{Md Sabir Ali\footnote{sabir.ali@iitrpr.ac.in},\, Sourav Bhattacharya\footnote{sbhatta@iitrpr.ac.in},\,\, Shankhadeep Chakrabortty\footnote{s.chakrabortty@iitrpr.ac.in}~~and~Shagun Kaushal\footnote{2018phz0006@iitrpr.ac.in}\\
\small{Department of Physics, Indian Institute of Technology Ropar, Rupnagar, Punjab 140 001, India}\\}
\begin{document}
\maketitle
\begin{abstract}
\noindent
 The violation of the Bell inequality for Dirac fermions is investigated in the cosmological de Sitter spacetime, in the presence of  background electromagnetic fields of constant strengths. The orthonormal Dirac mode functions are obtained and the relevant in-out squeezed state expansion in terms of the Bogoliubov coefficients are found. We focus on two scenarios here : strong electric field and heavy mass limits (with respect to the Hubble constant). Using the squeezed state expansion, we then demonstrate  the  Bell violations for the vacuum and some maximally entangled initial states.  Even though a background magnetic field alone cannot create particles, in the presence of background electric field and or spacetime curvature, it can affect the particle creation rate.  Our chief aim thus here is to investigate the role of the background magnetic field strength in the  Bell violation. Qualitative differences in this regard for different maximally entangled initial states are shown.  Further extension of these results to the so called  $\alpha$-vacua are also discussed. 
  \end{abstract}
\newpage
\tableofcontents
\section{Introduction}\label{S1}

One of the most outstanding features of quantum mechanics is certainly the entanglement, associated with the non-local properties of the quantum mechanical measurement procedure~\cite{bell_4, Bell, CHSH, Werner:1989zz, Vidal:1998re, bell_3, seprability, bell1:2003, bell_1}. After experimental confirmation, this has been placed on firm physical grounds~\cite{Aspect1, Aspect2}. We refer our reader to e.g.~\cite{NHB:1998zz, Mermin, Klyshko,
 Plenio:2007zz, NielsenChuang, Monogamy} and references therein for extensive  reviews and pedagogical discussions on quantum entanglement  and its various measures. 

A very important and useful measure of quantum entanglement is the violation of the Bell inequality~\cite{Bell, CHSH} (see also~\cite{NielsenChuang} for an excellent pedagogical discussion), which has been confirmed experimentally~\cite{Aspect1, Aspect2}. Such violation clearly rules out the so called classical hidden variable theories and establishes the probabilistic and (for entangled states) the non-local characteristics associated with the quantum measurement procedure, e.g.~\cite{bell_3, bell1:2003, bell_1} (also references therein). The Bell inequality was originally designed for bipartite pure states, which was later extended to multipartite systems, altogether known as the Bell-Mermin-Klyshko  inequalities (or the  Clauser-Horne-Shimony-Holt inequality),~\cite{CHSH, NHB:1998zz, Mermin, Klyshko}.

There are a couple of distinct relativistic sectors where entanglement properties of quantum fields  emerge very naturally, due to the creation of entangled particle pairs. The first is the maximally extended non-extremal black hole spacetimes, or the Rindler spacetime, where the entanglement of quantum fields between two causally disconnected spacetime wedges have been investigated, e.g.~\cite{BKAY:2017, FuentesSchuller:2004xp, degradation:2015, nper:2011, degradation:2015n}. The second is the cosmological backgrounds where the vacuum in the asymptotic future (the out vacuum) is related to that of the asymptotic past (the in vacuum) via squeezed state expansion, due to pair creation. We refer our reader to 
e.g.~\cite{EE, Maldacena:2015bha, bell:2017, Fuentes:2010dt, vaccum EE for fermions, SSS, Bhattacharya:2019zno, SHN:2020, Choudhury:2016cso, Choudhury:2017bou} and references therein for discussions on various measures of bosonic and fermionic fields in different coordinatisation of the de Sitter spacetime. Even in the flat spacetime particle pair creation is possible in the presence of a `sufficiently' strong background electric field,  viz the Schwinger pair creation, e.g.~\cite{Parker:2009uva}. Various aspects of entanglement properties between created particle-anti-particle pairs in the Schwinger mechanism including the effect of a background magnetic field can be seen in~\cite{Ebadi:2014ufa, Li:2016zyv, Agarwal:2016cir, Li:2018twv, Karabali:2019ucc, Dai:2019nzv, HSSS:2020}.  We also refer our reader to e.g.~\cite{Balasubramanian, Momentum space entanglement} for interesting aspects of entanglement in the flat space quantum field theory and to~\cite{Ryu:2006bv, Ryu:2006ef} for holographic aspects of entanglement. 

The study of entanglement in the context of the early inflationary era can give us insight about the state of a quantum field in the early universe.   Such investigations should not be regarded as mere academic interests, as attempts have been made to predict their possible observational signatures as well.  Specifically, entanglement generated in the early universe can affect the cosmological correlation functions or the cosmic microwave background (CMB). For example, the fermionic entanglement may lead to the breaking of scale invariance of  the inflationary power spectra~\cite{Boyanovsky:2018soy}. It was argued in~\cite{Rauch:2018rvx}
by studying the violation of the Bell inequality by the photons coming from certain high redshift quasars that they are entangled, indicating the existence of entangled quantum states in the early universe. We also refer our reader to~\cite{Morse:2020mdc} and references therein for discussion on signature of Bell violation in the CMB and its observational constraints  pertaining the Bell operators and some course graining parameter.

In the cosmological spacetimes particle pair creation occur due to the background spacetime curvature, e.g.~\cite{Parker:2009uva}.  However if background electromagnetic fields are also present there, the particle creation can  further be affected. A particularly interesting such scenario is the early inflationary spacetime endowed with primordial electromagnetic fields. Computations on the Schwinger effect for both bosonic and fermionic fields in the de Sitter spacetime and its possible connection to the observed magnetic field in the inter-galactic spaces (i.e., the so called galactic dynamo problem,~\cite{Subramanian:2015lua}), can be seen in e.g.~\cite{vilenkin, Xue:2017, Xue:2017cex, Xue:2017ecx, yoko:2016tty, Bavarsad:2017oyv}. 

In this paper we wish to compute the Bell violation for fermions in the cosmological de Sitter spacetime, in the presence of constant background electromagnetic fields. Previous studies on cosmological Bell violation can be seen in e.g.~\cite{Maldacena:2015bha, bell:2017, SHN:2020}.   Note that a magnetic field alone cannot create vacuum instability~\cite{Parker:2009uva}, can be intuitively understood as follows. Let us imagine a particle-antiparticle pair is created due  to the application of a magnetic field. The must move in opposite directions to get separated. However, the magnetic Lorentz force, $e \vec{v}\times \vec{B}$ acts in the same direction for both particle and anti-particle. Thus by applying a magnetic field alone, no matter how strong it is, we cannot create pairs. However, one may expect that in the presence of background spacetime curvature and or electric field, it can affect the pair creation rate. The entanglement will also certainly vary if the pair creation rate is altered.

 In a flat spacetime, pair creation only due to a background electric field is expected to cease upon the application of a magnetic field of sufficiently high strength, due to the aforementioned oppositely directed Lorentz force created by them.  Accordingly, the degradation of correlation or information between entangled states due to particle creation would also cease, as  has been shown recently in~\cite{HSSS:2020}. Let us now consider in addition, the spacetime curvature which would also create particle pairs.  Will the magnetic field be able to stop the particle creation  due to the gravitational field? The intuitive answer is No, as follows. In a pure gravitational background, a created particle pair will follow geodesics  and become observables in a spacetime like the de Sitter due to the geodesic deviation~\cite{Mironov:2011hp}. Such deviations  happen even for  initially parallel trajectories. Thus   as the particle-antiparticle pair created in the presence of geometric curvature  propagate, they are expected to get separated irrespective of the presence of Lorentz force imparted by the background magnetic field, even though that force is acting in the same direction for both of them.  This also indicates that in the absence of an electric field, the magnetic field perhaps cannot affect the particle creation due to the gravitational field at all. We shall check  these intuitive guesses explicitly in the next section. Our goal here is to study the effect of the background magnetic field strength on the Bell violation.

Apart from this, a physical motivation behind this study comes from the possible connection between the primordial electromagnetic fields and the  aforementioned galactic dynamo problem, e.g.~\cite{Xue:2017}. We wish to consider fermions instead of a complex scalar, as the former are more realisitic.  Let us  speculate about some possible observational consequences of the model we study. For example, one can compute the power spectra by tracing out the fermionic degrees of freedom (interacting with the inflaton or gravitational excitations) and check the breaking of  scale invariance  as of~\cite{Boyanovsky:2018soy}. Likewise if we also consider the quantum part of the electromagnetic sector, it should carry information about the entangled fermionic states once we trace out the fermionic degrees of freedom, originating from the photon-fermion interaction.  Thus one can expect that the photons coming from the distant past undergoing the Bell test as of~\cite{Rauch:2018rvx}, will carry information about such entangled fermionic states. Since these states are defined  in the presence of the primordial background electromagnetic fields, the Bell test might also carry information about those background fields. This can possibly be used to constrain the corresponding  field strengths and test  the proposition 
of~\cite{Xue:2017}.  With this motivation, and as a problem to begin with, we shall simply compute below the fermionic  Bell violation in the cosmological de Sitter spacetime in the presence of background electromagnetic fields, as a viable measure of quantum entanglement.

The rest of the paper is organised as follows. In \ref{S2} and \ref{A}, we compute the orthonormal in  and out Dirac modes in the cosmological de Sitter spacetime in the presence of constant background electric and magnetic fields. The Bogoliubov coefficients and the squeezed state relationship between the in and out vacua are also found. Using this, we compute the vacuum entanglement entropy in~\ref{EE}. The Bell inequality violations for the vacuum and also  two maximally entangled initial states are computed in \ref{BV}. All these results are further extended in \ref{alph} to the so called one parameter fermionic $\alpha$-vacua. Finally we conclude in \ref{sec:SD}. We shall assume  either the field is  heavily massive or the electric  field strength is very high (with respect to the Hubble constant).

We shall work with the mostly positive signature of the metric in $(3+1)$-dimensions and  will set $c=1=\hbar$ throughout.

\section{The in and out Dirac  modes}\label{S2}

For our purpose, we first need to solve the Dirac equation in the cosmological de Sitter spacetime in the presence of constant background electromagnetic fields. The following will be an extension of the solutions found earlier in the same spacetime but in the absence of any magnetic field~\cite{Xue:2017cex, yoko:2016tty}.

The Dirac equation in a general curved spacetime reads,
\begin{equation}
\label{diraceqincurve}
 (i\gamma^{\mu}D_{\mu}-m)\psi(x)=0
\end{equation}
where  the gauge cum spin covariant derivative reads, 
$$D_\mu\equiv \partial_\mu+ieA_\mu+\Gamma_\mu$$
The spin connection is given by,
\begin{eqnarray}
\label{conn1}
\Gamma_\mu=-\frac{1}{8}e^{\mu}_a \left(\partial_{\mu}e_{b\nu}- \Gamma_{\mu\nu}^{\lambda} e_{b\lambda}\right)[\gamma^a, \gamma^b],
\end{eqnarray}
where the latin indices represent the local inertial frame and  $e^{\mu}_a$ are the tetrads.

The de Sitter metric in $(3+1)$-dimensions reads,
\begin{eqnarray}
\label{FLRW}
ds^2=\frac{1}{H^2\eta^2}\left(-d\eta^2+dx^2+dy^2+dz^2\right),
\end{eqnarray}
where $H$ is the Hubble constant  and the conformal time $\eta$ varies from $-\infty <\eta < 0^-$. Choosing now
$e^{a}_\mu=a(\eta)\delta^a_\mu$, we have from \ref{conn1},
\begin{equation}
\Gamma_{\mu}=\frac{1}{2}\gamma^{\mu}\gamma^0 a^\prime(\eta) a(\eta) \delta_{\mu}^i,\;\;i=1,2,3,    
\label{conn2}
\end{equation}
where the prime denotes differentiation once with respect to  $\eta$.

Defining a new variable in terms of the scale factor $a(\eta)= -1/H\eta$, as
\be \label{conf}\xi=a^{\frac32}\psi,\ee
and using \ref{conn2}, the Dirac equation \ref{diraceqincurve}  becomes,
\begin{eqnarray}
\label{dirac_3}
\left(ie^\mu_a  \gamma^a \partial_\mu-eA_\mu e^\mu_a \gamma^a-m\right)\xi(\eta,\vec{x})\;=\;0.
\end{eqnarray}
Substituting next
\be
\xi(\eta,\vec{x})=\left(ie^\mu_a \gamma^a\partial_\mu-eA_\mu e^\mu_a \gamma^a+m\right)\zeta(\eta,\vec{x})
\label{de0}
\ee
 into \ref{dirac_3}, we obtain the squared Dirac equation
\begin{eqnarray}
\label{sq_dirac}
\left[\left(\partial_\mu+ieA_\mu\right)^2-m^2a^2+i\left(ma^\prime a e^0_0 \gamma^0-\frac{e}{2}a^2 e^\mu_a \gamma^ae^\nu_a \gamma^b F_{\mu \nu}\right)\right]\zeta(\eta,\vec{x})\;=\;0
\end{eqnarray}
We choose the gauge  to obtain constant electric and magnetic fields in the $z$-direction as, 
\begin{eqnarray}
\label{pot1}
A_\mu=By\delta^{x}_\mu-\frac{E}{H}\left(a-1\right)\delta^z_\mu,
\end{eqnarray}
where $E$, $B$ are constants. Making now the ansatz  
$$\zeta(\eta,\vec{x})=e^{-iez E/H}e^{{i\vec{k}_\slashed{y}}\cdot\vec{x}}\zeta_{s}(\eta,y)\omega_s,$$
in \ref{sq_dirac}, where $\vec{k}_\slashed{y}=(k_x,0,k_z)$, we have
\begin{eqnarray}
\label{dirac4}
&&\Bigg[\left(\partial_y^2-\left(k_x+eBy\right)^2\right)-\partial_0^2-k_z^2+2HLak_z-H^2L^2a^2-m^2a^2+i Ha^2\left(M\gamma^0+L\gamma^0\gamma^3 +\frac{eB}{H a^2}\gamma^1\gamma^2\right)\Bigg]\zeta_{s}(\eta,y)\omega_s=0,\nonumber\\
\end{eqnarray}
where 
$$M=\frac{m}{H} \qquad {\rm and} \qquad  L=\frac{eE}{H^2},$$
 are dimensionless mass and electric field strengths. Note also in \ref{dirac4} that the matrices $(M\gamma^0+L\gamma^0\gamma^3) $ and $\gamma^1\gamma^2$ commute and hence we may treat $\omega_s$ to be their simultaneous eigenvectors. Thus \ref{dirac4} becomes
\begin{eqnarray}
\label{sq_dirac2}
\left[\left(\partial_y^2-\left(k_x+eBy\right)^2\right)+\left(-\partial_0^2-\omega_k^2+i\lambda_s\sigma(\eta)+ie\beta_s B\right)\right]\zeta_{s}(\eta,y) =0,
\end{eqnarray} 
where $\lambda_s=\pm 1$, $\beta_s=\pm i$ and  we have abbreviated,
\begin{eqnarray}
\label{const}
\omega_k^2=k_z^2-2HLak_z+H^2a^2\left(L^2+M^2\right),\;\sigma(\eta)=a^2H^2\sqrt{L^2+M^2}.
\end{eqnarray} 
The explicit expressions for the four orthonormal eigenvectors $\omega_s$ are given in \ref{A}.
Substituting now for the variable separation, $\zeta_{s}(\eta,y)=\varsigma_s(\eta) h_s(y)$ into \ref{sq_dirac2}, we obtain the decoupled equations,
\begin{eqnarray}
\left(\partial_\eta^2+\omega_k^2-i\lambda_s\sigma(\eta)+S_s\right){\varsigma_{s}(\eta)}=0,  \qquad {\rm and }\qquad \left(\partial_y^2-\left(k_x+eBy\right)^2+S_s+i\beta_s eB \right) h_s(y)=0,
\label{dirac9}
\end{eqnarray}
where $S_s$ is the separation constant. Clearly, we can have four sets of such pair of equations corresponding to the different choices of $\lambda_s =\pm 1,\,\, \beta_s = \pm i$. For example, for $\lambda_s=1$, $\beta_s =-i$ and 
$\lambda_s=1$, $\beta_s =i$, we respectively have,
\begin{eqnarray}
&&\left(\partial_\eta^2+\omega_k^2-i\sigma(\eta)+S_1\right){\varsigma_{1}(\eta)}=0,  \qquad {\rm and }\qquad \left(\partial_y^2-\left(k_x+eBy\right)^2+S_1+ eB \right) h_1(y)=0, \nonumber\\
&& \left(\partial_\eta^2+\omega_k^2-i\sigma(\eta)+S_2\right){\varsigma_{2}(\eta)}=0,  \qquad {\rm and }\qquad \left(\partial_y^2-\left(k_x+eBy\right)^2+S_2- eB \right) h_2(y)=0, 
\label{dirac9'}
\end{eqnarray}

Let us first focus on the spatial equations. In terms of the variable
 $$\overline{y}=\left(\sqrt{eB}y+\frac{k_x}{\sqrt{eB}}\right),$$
 it is easy to see that the spatial differential equations of \ref{dirac9'} reduce to the Hermite differential equation, with the separation constants,
 $$S_1=2n eB \quad \text{and} \quad S_2=2(n+1)eB,$$ 
 where $n=0,1,2...$ denote  the Landau levels. Thus we have the normalised solutions,
$$h_1(y)=h_2(y)=\left(\frac{\sqrt{eB}}{2^{n+1}\sqrt{\pi}(n+1)!}\right)^{1/2}e^{-\overline{y}^2/2}\mathcal{H}_{n}(\overline{y})=h_{n}(\overline{y})~({\rm say}),$$
where $\mathcal{H}_{n}(\overline{y})$ are the Hermite polynomials of order $n$.  \\

For the two temporal equations in \ref{dirac9'}, we introduce the variables,
$$z_1=-\frac{2i\sqrt{k_z^2+S_1}}{aH}\quad \text{and}\quad  z_2=-\frac{2i\sqrt{k_z^2+S_2}}{aH},$$ 
so that they respectively become,
\begin{eqnarray}
\label{zcom1}
\left(\partial^2_{z_{1}}-\frac{1}{4}+\frac{\kappa_1}{z_1}+\frac{(1/4-\mu^2)}{z_1^2}\right){\varsigma_1(z_1)}=0\qquad {\rm and} \qquad 
\left(\partial^2_{z_{2}}-\frac{1}{4}+\frac{\kappa_2}{z_2}+\frac{(1/4-\mu^2)}{z_2^2}\right){\varsigma_2(z_2)}=0,
\end{eqnarray}
where we have abbreviated,
\begin{eqnarray}
\label{coeff}
\kappa_1=-\frac{ik_z L}{\sqrt{k_z^2+S_1}}, \qquad
\kappa_2=-\frac{ik_z L}{\sqrt{k_z^2+S_2}}, \qquad 
\mu=\left(\frac{1}{2}+i\sqrt{M^2+L^2}\right).
\end{eqnarray}
Note that $\kappa_{1,2}$ depend upon the sign of $k_z$. From now on we shall only focus on the situation $(M^2+L^2)\gg 1$, for which 
$$\mu\approx i\sqrt{M^2+L^2} \approx i |\mu|,$$
 in \ref{coeff}.  This corresponds to either very  strong electric field or  highly massive field, or both.   Then the general solutions for \ref{zcom1} are given by,
\begin{eqnarray}
\label{sols1}
{\varsigma_1(z_1)}=C_1 W_{\kappa_1,i|\mu|}(z_1)+D_1M_{\kappa_1,i|\mu|}(z_1),\qquad {\rm and }\qquad 
{\varsigma_2(z_2)}=C_2 W_{\kappa_2,i|\mu|}(z_2)+D_2M_{\kappa_2,i|\mu|}(z_2).
\end{eqnarray}
where $W$ and $M$ are the Whittaker functions~\cite{AS} and $C_1,\,C_2,\,D_1,\,D_2$ are constants.

Let us now find out the positive frequency `in' modes, i.e. the mode functions whose temporal part behaves as positive frequency plane waves as $\eta \to -\infty$. In this limit we have~\cite{AS},
$$W_{\kappa_1,i|\mu|}(z_1)\sim e^{-2i\eta \sqrt{k_z^2+S_1}}\eta^{\kappa_1}$$
Thus for such modes we must set $D_1=0=D_2$ in \ref{sols1}. Putting things together, we write the two positive frequency in mode functions as,
\begin{eqnarray}
\label{in_comp}
\zeta^{\text{in}}_{s,n}({\eta,\vec{x}})&=& e^{-iHLz}e^{i\vec{k}\slashed{y}\cdot\vec{x}}W_{\kappa_s,i|\mu|}(z_s)h_{n}(\overline{y})\,\omega_s \qquad (s=1,2)
\end{eqnarray}

Likewise, since as $\eta \to 0^-$~\cite{AS}
$$M_{\kappa,i|\mu|}(z_1)\sim \eta^{i|\mu|+1/2},$$ 
the positive frequency out modes can be defined with respect to the cosmological time, $t$ ($t=-\ln H\eta/H $) and we choose them to be 
\begin{eqnarray}
\label{out_comp}
\zeta^{\text{out}}_{s,n}({\eta,\;\vec{x}})= e^{-iHLz}e^{i\vec{k}\slashed{y}\cdot\vec{x}}M_{\kappa_s,i|\mu|}(z_s)h_{n}(\overline{y})\,\omega_s \qquad (s=1,2)
\end{eqnarray}

However, recall that the $\zeta$'s appearing in \ref{in_comp} and \ref{out_comp} are not the original Dirac modes, as of  \ref{conf},  \ref{sq_dirac}. We thus have the complete set of positive and  negative frequency in and out modes,
\begin{eqnarray}
\label{mainmodes}
U_{s,n}^{\rm in}&=&\frac{1}{N_s a^{3/2}} \hat{D}\zeta^{\text{in}}_{s,n},\qquad V_{s,n}^{\rm in}\;=\; \mathcal{C}(U_{s,n}^{\rm in})^*,\nonumber\\
U_{s,n}^{\rm out}&=&\frac{1}{M_s a^{3/2}}\hat{D}\zeta^{\text{out}}_{s,n},\qquad V_{s,n}^{\rm out}\;=\;\mathcal{C}(U_{s,n}^{\rm out})^*, \qquad (s=1,2)
\end{eqnarray}
where $\mathcal{C}= i \gamma^2$ is the charge conjugation matrix. Hence the $V$-modes appearing above are the negative frequency modes. The normalisation constants appearing above are given by
\be N_1 = e^{\pi |\kappa_1|{\rm sgn}(k_z)/2}, \qquad N_2 = e^{\pi |\kappa_2|{\rm sgn}(k_z)/2}, \qquad M_1=M_2= \sqrt{2|\mu|} e^{\pi |\mu|/2}
\label{nc}
\ee
where the sign dependence of the normalisation constants originates from the sign dependence (of $k_z$) of the parameters $\kappa_s$, \ref{coeff}. 
The explicit form of the mode functions in \ref{mainmodes} and the evaluation of the normalisation constants are discussed in \ref{A}.

It is easy to check that these mode functions satisfy the orthonormality relations,
\begin{eqnarray}
&&(U_{s,n}^{\rm in}(x;\vec{k}_{\slashed{y}}),U_{s^{\prime},n^{\prime}}^{in}(x;\vec{k^{\prime}_{\slashed{y}}}))= (V_{s,n}^{\rm in}(x;\vec{k}_{\slashed{y}}),V_{s^,n^{\prime}}^{\rm in}(x;\vec{k^{\prime}_{\slashed{y}}})) = \delta^{2}(\vec{k}_{\slashed{y}}-\vec{k'_{\slashed{y}}})\delta_{n n^{\prime}}\delta_{ss^{\prime}}\nonumber\\
&&(U_{s,n}^{\rm out}(x;\vec{k}_{\slashed{y}}),U_{s^{\prime},n^{\prime}}^{\rm out}(x;\vec{k^{\prime}_{\slashed{y}}}))= (V_{s,n}^{\rm out}(x;\vec{k}_{\slashed{y}}),V_{s^{\prime},n^{\prime}}^{\rm out}(x;\vec{k^{\prime}_{\slashed{y}}})) = \delta^{2}(\vec{k}_{\slashed{y}}-\vec{k'_{\slashed{y}}})\delta_{nn^{\prime}}\delta_{s,s^{\prime}},
\end{eqnarray}
with all the other inner products vanishing.

In terms of these orthonormal modes, we now make the field quantisation, 
\begin{eqnarray}
\psi(\eta,\vec{x}) &&= \sum_{n; s=1,2}\int\frac{d^{2}\vec{k}_{\slashed{y}}}{2\pi a^{3/2}} \Bigg[a_{\rm in}(\vec{k}_{\slashed{y}},s,n)U_{s,n}^{\rm in}(x;\vec{k}_{\slashed{y}})+b^{\dagger}_{\rm in}(\vec{k}_{\slashed{y}},s,n)V_{s,n}^{\rm in}(x;\vec{k}_{\slashed{y}})\Bigg] \nonumber\\
&&=\sum_{n; s=1,2}\int\frac{d^{2}\vec{k}_{\slashed{y}}}{2\pi a^{3/2}}\sum_{n; s=1,2}\Bigg[a_{\rm out}(\vec{k}_{\slashed{y}},s,n)U_{s,n}^{\rm out}(x;\vec{k}_{\slashed{y}})+b^{\dagger}_{\rm out}(\vec{k}_{\slashed{y}},s,n)V_{s,n}^{\rm out}(x;\vec{k}_{\slashed{y}})\Bigg],
\label{field}
\end{eqnarray}
where the creation and annihilation operators are assumed to satisfy the usual canonical anti-commutation relations.\\

Using now the relations between the Whittaker functions~\cite{AS}, 
\begin{eqnarray}
\label{Whitidentity}
  && W_{\kappa,i|\mu|}(z) = \frac{\Gamma(-2i|\mu|)}{\Gamma(1/2-i|\mu|-\kappa)}M_{\kappa,i|\mu|}(z) + \frac{\Gamma(2i|\mu|)}{\Gamma(1/2+i|\mu|-\kappa)}M_{\kappa,-i|\mu|}(z)\nonumber\\
 &&   M_{\kappa,i|\mu|}(z)= 
    -i e^{\pi |\mu|} M_{-\kappa,i|\mu|}(-z)
\end{eqnarray}
into \ref{in_comp}, \ref{out_comp} and \ref{mainmodes},   we find
\begin{equation}
\label{bogo1}
U_{s,n}^{\rm in}(x;\vec{k}_{\slashed{y}})= \frac{M_s}{N_s} \frac{\Gamma(-2i|\mu|)}{\Gamma(1/2-i|\mu|-\kappa_s)} U_{s,n}^{\rm out}(x;\vec{k}_{\slashed{y}}) + i e^{-|\mu|\pi}\frac{M_s}{N_s}  \frac{\Gamma(2i|\mu|)}{\Gamma(1/2+i|\mu|-\kappa_s)} V_{s,n}^{\rm out}(x;\vec{k}_{\slashed{y}}) \qquad (s=1,2)
\end{equation}
Substituting this into \ref{field}, we obtain the Bogoliubov relations
\begin{eqnarray}
\label{in_out}
&&a_{\rm out}(\vec{k}_{\slashed{y}},s,n)\;=\;\alpha_s a_{\rm in}(\vec{k}_{\slashed{y}},s,n)-\beta_{s}^{*} b_{\rm in}^{\dagger}(-\vec{k}_{\slashed{y}},s,n)\nonumber\\
&&b_{\rm out}(\vec{k}_{\slashed{y}},s,n)\;=\;\alpha_s b_{\rm in}(\vec{k}_{\slashed{y}},s,n)+\beta_{s}^{*} a^{\dagger}_{\rm in}(-\vec{k}_{\slashed{y}},s,n)
\end{eqnarray}
The canonical anti-commutation relations ensure,
$$|\alpha_s|^2+|\beta_s|^2=1 \qquad (s=1,2)$$
Recalling we are working with $|\mu|\gg1 $, we have 
\be
\label{complxbv}
\beta_s\approx  i e^{-|\mu|\pi}\left\vert\frac{M_s}{N_s}\right\vert   \frac{\Gamma(2i|\mu|)}{\Gamma(i|\mu|-\kappa_s)} \qquad (s=1,2)
\ee
We find from the above after using some identities of the gamma function~\cite{AS}, the spectra of pair creation 
\begin{equation}
\label{number_demsity}
|\beta_s|^2_{\pm}=e^{-\pi(|\mu|\pm |\kappa_s|)}\frac{\sinh\pi(|\mu|\pm|\kappa_s|)}{\sinh2\pi |\mu|}\qquad (s=1,2)
\end{equation}
where the $\pm$ sign correspond respectively to $k_z > 0$ and  $k_z < 0$, originates from the fact that $\kappa_s$ depend upon the sign of $k_z$, \ref{coeff}. The above expression is formally similar to the case where only a background electric field  is present \cite{Xue:2017cex}. 
The contribution to the particle creation 
from the magnetic field comes solely from the coefficients $\kappa_s$ and there is no contribution of it (i.e., $\kappa_s=0$) if either the electric field is vanishing, Or, the magnetic field strength is infinitely large. Note  also  that if we set $E=0$ in \ref{number_demsity}, we reproduce the well known fermionic blackbody spectra of created particles  with temperature $T_H=H/2\pi$, e.g.~\cite{SHN:2020},
\be
|\beta_s|^2_{\pm}=\frac{1}{e^{2\pi |\mu|}+1}\qquad (s=1,2)
\ee
where $|\mu| =M=m/H$. The above discussions show  that in the absence of an electric field, the magnetic field cannot alter the particle creation rate, as we intuitively anticipated towards the middle of \ref{S1}. Finally, we also note from \ref{number_demsity} that since $|\mu|=(M^2+L^2)^{1/2}$, for $E\neq 0$, and even if $B \to \infty$, the particle creation due to the electric field does not completely vanish, unlike that of the flat spacetime~\cite{HSSS:2020}. Once again, this should correspond to the fact that the mutual separation of the pairs created  by the electric field as they propagate, is also happening here due to the expanding gravitational field of the de Sitter, upon which the magnetic field has no effect. \\

Since the parameters $M$ and $L$ denote the dimensionless rest mass and the strength of the electric field  (cf., discussion below \ref{dirac4}), let us consider in the following two qualitatively distinct cases, keeping in mind $(M^2+L^2)\gg 1$.\\

\noindent
\textbf{Case $1$}: $M^2\gg1\quad {\rm and}\quad  M^2\gg L^2$.  Hence in this case particle creation is happening chiefly due to the background spacetime curvature. We have from \ref{number_demsity} in this limit,
\begin{equation}
	\label{case1}
	|\beta_s|^2_{\pm} \approx e^{-2 \pi M}\left(1- e^{- 2 \pi M}e^{\mp 2\pi|\kappa_s|}+{\cal O}(e^{-4\pi M})\right) \,	\end{equation}
Thus $|\beta_s|^2_+> |\beta_s|^2_-$. Although the electromagnetic field is weak here and hence they would have little effects on the particle creation, note in particular from the expression of $\kappa_s$ that if we keep the electric field strength and $k_z$ fixed,  $|\beta_s|^2_+$ decreases whereas $|\beta_s|^2_-$ increases with the magnetic field strength, and for extremely high $B$-value, the particle creation rate coincides to that of only due to the spacetime curvature. \\

\noindent
	\textbf{Case $2$}: $L^2\gg1\quad {\rm and}\quad  L^2 \gg M^2$. Hence in this case particle creation is happening chiefly due to the background electric field. We have from \ref{number_demsity},
	\begin{equation}
	\label{case2}
	|\beta_s|^2_{\pm} \approx e^{-2 \pi L} \left(1-e^{-2 \pi L}e^{\mp 2 \pi |\kappa_s|}+ {\cal O}(e^{-4\pi L})\right) \,	\end{equation}
Thus in this case  also $|\beta_s|^2_+>|\beta_s|^2_-$, and $|\beta_s|^2_+$ decreases whereas $|\beta_s|^2_-$ increases with the magnetic field strength, while the other parameters are held fixed. We wish to focus only on   $|\beta_s|^2_-$ in the following. In our computation, we shall often encounter  the complex $\beta_{s}$ value. Hence instead of using \ref{case1} or \ref{case2}, we shall work with  \ref{complxbv}, by taking numerical values of the parameters  appropriate for the particular case.

\bigskip

Subject to the field quantisation in \ref{field}, the in and the out vacua are defined as,
$$a_{\rm in}(\vec{k}_{\slashed{y}},s,n) |0\rangle^{\rm in}=0 = b_{\rm in}(\vec{k}_{\slashed{y}},s,n) |0\rangle^{\rm in}\qquad {\rm and }\qquad a_{\rm out}(\vec{k}_{\slashed{y}},s,n) |0\rangle^{\rm out}=0 = b_{\rm out}(\vec{k}_{\slashed{y}},s,n) |0\rangle^{\rm out}$$

Thus the Bogoliubov relations  \ref{in_out} imply a (normalised) squeezed state expansion between the in and out states for a given momentum, 
\begin{eqnarray}
\label{vaccum1}
|0_{k}\rangle^{\rm in}\;
= \left(\alpha_1 |0_{k}^{(1)}\,0_{-k}^{(1)}\rangle^{\rm out} +\beta_{1} |1_{k}^{(1)}\,1_{-k}^{(1)}\rangle^{\rm out}\right )\otimes  \left(  \alpha_2 |0_{k}^{(2)}\,0_{-k}^{(2)}\rangle^{\rm out} +\beta_{2} |1_{k}^{(2)}\,1_{-k}^{(2)}\rangle ^{\rm out}\right)
\end{eqnarray}
As we have discussed above, since we shall be working only with the `$-$' sign of \ref{number_demsity}, $\beta_s$ and $\alpha_s$ appearing above are understood as  $\beta_{1-}$, $\beta_{2-}$ and $\alpha_{1-}$ and $\alpha_{2-}$, respectively.

The excited in states can be written in terms of the out states by applying the in-creation operators on the left hand side of \ref{vaccum1}, and then using the Bogoliubov relations  \ref{in_out} on its right hand side.

Finally we note that  the $s=1,2$ sectors are factorised in \ref{vaccum1}, leading to 
$$|0_{k}\rangle^{\rm in}  = |0_{k}^{(1)}\rangle^{\rm in}\otimes |0_{k}^{(2)}\rangle^{\rm in}$$
Thus for simplicity, we can work only with a single sector, say $|0_{k}^{(1)}\rangle^{\rm in}$, of the in-vacuum. 

Being equipped with these, we are now ready to go to the computation of the Bell violation. However before we do that, we wish to compute the entanglement entropy associated with the vacuum state.

\section{Entanglement entropy of the vacuum}
\label{EE}
\noindent
If a system is made of two subsystems say $A$ and $B$, 
the entanglement entropy of $A$ is defined as the von Neumann entropy of $\rho_A$,
$ S(\rho_A)=-	\text{Tr}_A\left(\rho_A\ln \rho_A\right)$, where $\rho_A$ is the reduced density operator, $\rho_A= \text{Tr}_B \rho_{AB}$. The entanglement entropy of $B$ is defined in a likewise manner. If $\rho_{AB}$ corresponds to a pure state, one has $S(\rho_A)=S(\rho_B)$, and it is vanishing when $\rho_{AB}$ is also separable,  $\rho_{AB}=\rho_A \otimes \rho_B  $. The von Neumann entropies satisfy a subadditivity, $S(\rho_{AB}) \leq 	S(\rho_A) + S(\rho_B)$, where $S(\rho_{AB})$ is the Von Neumann entropy corresponding to $\rho_{AB}$, and the equality holds if and only if $\rho_{AB}$ is separable, e.g.~\cite{NielsenChuang}.\\

We wish to compute the entanglement entropy for the state $|0_{k}^{(1)}\rangle^{\rm in}$, defined at the end of the preceding section.  The density matrix  corresponding to this state is pure, $\rho_0= |0_{k}^{(1)}\rangle^{\rm in\,\,in}\langle0_{k}^{(1)}|$. Using \ref{vaccum1}, we write down $\rho_0$ in terms of the out states, which contain both $k$ and $-k$ degrees of freedom. The reduced density matrix corresponding to the $k$ sector (say, particle) is given by, 
$
	\rho_k
=
	\text{Tr}_{-k}
	\rho_0
=
	|\alpha_1|^{2} |0_{k}^{(1)}\rangle^{\rm out\,\, out}\langle 0_{k}^{(1)}| + |\beta_1|^{2} |1_{k}^{(1)}\rangle^{\rm out\,\, out}\langle 1_{k}^{(1)}|
$, and hence the entanglement entropy
is give by 
\begin{equation}
    \label{entropy}
	S_k =
	-
	\text{Tr}_k
	\rho_k
	\ln \rho_k
=
	-   \left[ \ln	(1-|\beta_1|^{2})    +|\beta_1|^2  \ln \frac{|\beta_1|^{2}}{1-|\beta_1|^2}\right]
\end{equation}
We also have $S_k=S_{-k}$, as we are dealing with a pure state.

\begin{figure}[h]
	\centering
		\includegraphics[scale=.53]{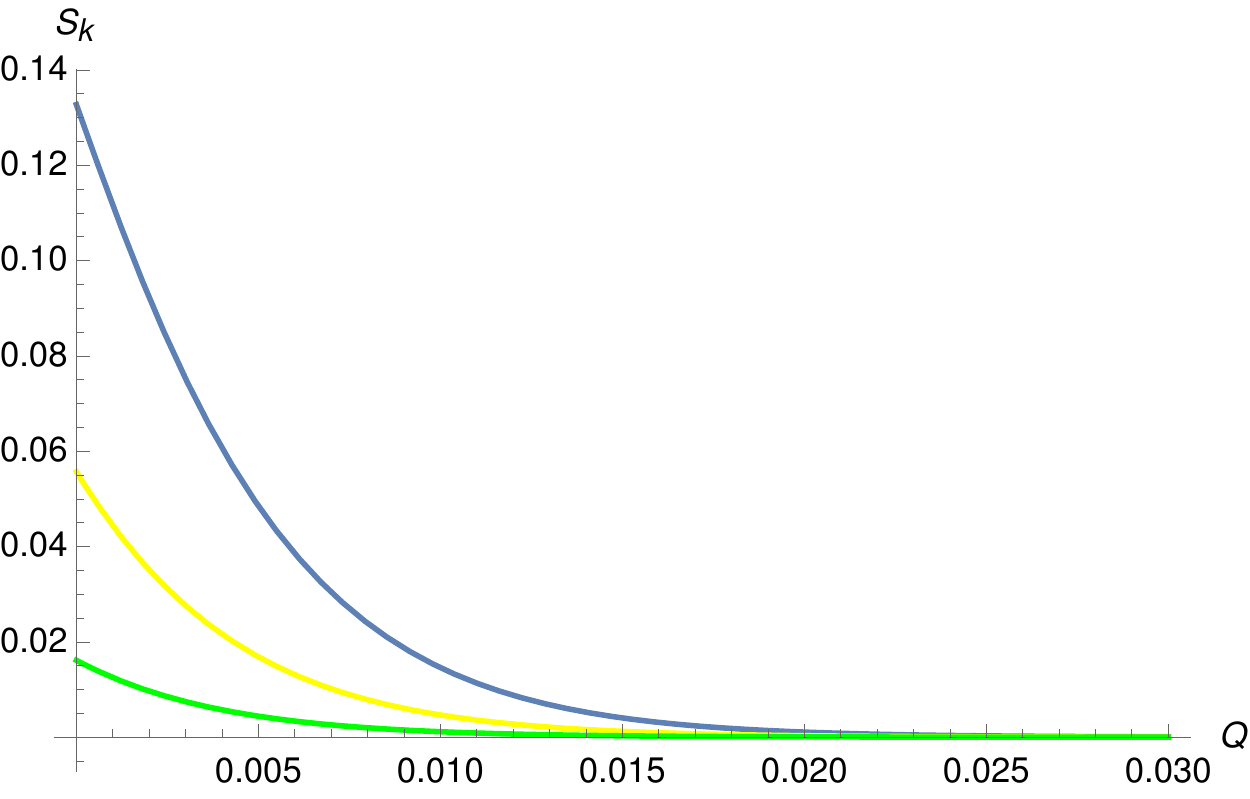}\hspace{1.0cm}
		\includegraphics[scale=.53]{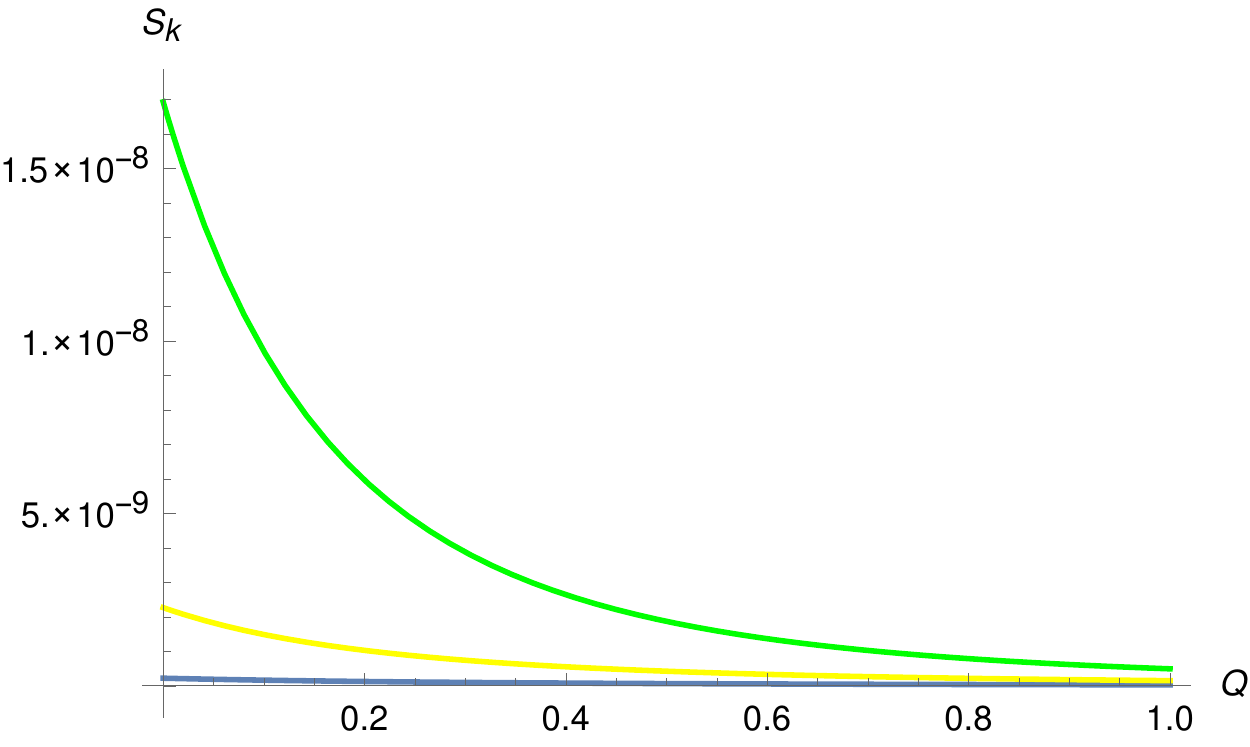} 
		\caption{We have plotted entanglement entropy \ref{entropy} corresponding to the vacuum state with respect to the parameter $Q=2neB/k^2_z$. The left plot corresponds to the case $L^2\gg M^2$, where we have taken $L=100$ and different curves correspond to different $M$ values (blue $M=10$, yellow $M=12$ and green $M=16$). The right plot corresponds to $M^2\gg L^2$, where we have taken $M=5$, and different curves correspond to different $L$ values (blue $L=1$, yellow $L=1.5$ and green $L=2$). For a given mode, the entanglement entropy decreases monotonically with increasing $B$ for both the cases due to the decrease in the particle creation. See main text for detail. 
		}
		\label{fig:EEforVacuum1}
\end{figure}

Since we are chiefly interested here on the effect of the magnetic field strength, let us extract a dimensionless parameter from \ref{coeff},
$$Q=\frac{2enB}{k^2_z}$$
The $Q$-dependence of $S_k$ is depicted in \ref{fig:EEforVacuum1}, for the two cases (`large' ($L^2\gg M^2$) and `small' ($L^2 \ll M^2$) electric fields), discussed in the preceding section. For a given mode (i.e. $n,\,k_z$ fixed) thus, the increase in $Q$ corresponds to the increase in the $B$ value. As can be seen in the figure, the entanglement entropy decreases monotonically with the increase in the magnetic field strength. This corresponds to the fact that the vacuum entanglement entropy originates from the pair creation, which decreases  with increasing $B$ for both the cases we have considered. 
%

\section{The violation of the Bell  inequalities}\label{BV}
\subsection{The Bell inequalities}
The construction of the Bell or the Bell-Mermin-Klyshko (BMK) operators for fermions are similar to that of the scalar field, e.g.~\cite{NielsenChuang, bell:2017} and references therein. Let us consider two pairs of non-commuting observables defined respectively over the Hilbert spaces ${\cal H}_{A}$ and  ${\cal H}_{B}$ : $({\mathit{O}_{1},\mathit{O}'_{1})\in{\cal H}_{A}} \; {\rm and}\; ({\mathit{O}_{2},\mathit{O}'_{2})\in {\cal H}_{B}} $. We assume that these are spin-$1/2$ operators along specific directions, such as $\mathit{O}=n_{i}\sigma_{i}$, $\mathit{O}'=n_{i}^{\prime}\sigma_{i}$, where $\sigma_{i}$'s are the Pauli matrices and $n_{i}$, $n_{i}^{\prime}$ are unit vectors on the three dimensional  Euclidean space. The eigenvalues of each of these operators are $\pm1$. The Bell operator,  ${\cal B}\in{\cal H}_{A}\otimes{\cal H}_{B}$ is defined as (suppressing the tensor product sign),
\begin{eqnarray}
\mathcal{B}=\mathit{O}_{1}\left(\mathit{O}_{2}+\mathit{O}'_{2}\right) +\mathit{O}'_{1}\left(\mathit{O}_{2}-\mathit{O}'_{2}\right)
\label{op2}
\end{eqnarray}
In theories with  classical local hidden variables, we have the so called Bell's inequality, $\langle\mathcal{B}^{2} \rangle \leq4$ and $|\langle\mathcal{B}\rangle|\leq2$. However, this inequality is violated in quantum mechanics as follows.  We have from \ref{op2},
\begin{eqnarray}
\mathcal{B}^{2} = {\bf I} - [\mathit{O}_{1},\mathit{O}'_{1}][\mathit{O}_{2},\mathit{O}'_{2}],
\end{eqnarray}
where {\bf I} is the identity operator. Using the commutation relations for the Pauli matrices one gets $|\langle \mathcal{B}\rangle| \leq\;2\sqrt{2}$, thereby obtaining a violation of Bell's inequality, where the equality is regarded as the maximum violation. 

The above construction can be extended to multipartite systems with pure density matrices as well, corresponding to squeezed states formed by mixing different modes. We refer our reader to~\cite{bell:2017} and references  therein for details.

We wish to investigate below Bell's inequality violation for the vacuum as well as some maximally entangled initial states. 

\subsection{Bell violation for the vacuum state}

We wish to find out the expectation value of $\mathcal{B}$, \ref{op2}, with respect to the vacuum state $|0_{k}^{(1)} \rangle^{\rm in}$, given at the end of \ref{S2}. In order to do this, one usually introduces pseudospin operators measuring the parity in the Hilbert space along different axes, e.g~\cite{bell:2017} and references therein. These operators for fermionic systems with eigenvalues $\pm 1$ are defined as,
\begin{equation}
\label{unitvector}
\mathbf{\hat{n}}.\mathbf{S} = S_{z}\cos\theta + \sin\theta(e^{i\phi}S_{-}+e^{-i\phi}S_{+}), 
\end{equation}
where $\mathbf{\hat{n}} = (\sin\theta \cos\phi,\sin\theta \sin\phi,\cos\phi)$ is a unit vector in the Euclidean $3$-plane. The action of the operators $S_z$ and $S_{\pm}$ are defined on the {\it out states},
\begin{equation}
\label{spin}
S_z | 0\rangle = - | 0\rangle, \quad S_z | 1\rangle =  | 1\rangle, \quad S_+| 0\rangle =| 1\rangle, \quad S_+| 1\rangle =0, \quad S_-| 0\rangle =0, \quad S_-| 1\rangle =|0\rangle
\end{equation}
Without any loss of generality, we take the operators to be confined to the $x-z$ plane, so that we may set  $\phi = 0 $ in \ref{unitvector}. We may then take in \ref{op2}, $\mathit{O_{i}}=\;\mathbf{\hat{n}}_i\cdot\mathbf{S}$  and $\mathit{O}'_i=\mathbf{\hat{n}'}_i\cdot\mathbf{S}$ 
with $i=1,2$. Here $\mathbf{\hat{n}}_i$ and $\mathbf{\hat{n}'}_i$ are two pairs of unit vectors in the Euclidean $3$-plane, characterised by their angles with the $z$-axis, $\theta_{i}$, $\theta'_{i}$ (with $i=1,2$) respectively. 

\begin{figure}[h]
\centering
\includegraphics[scale=0.60]{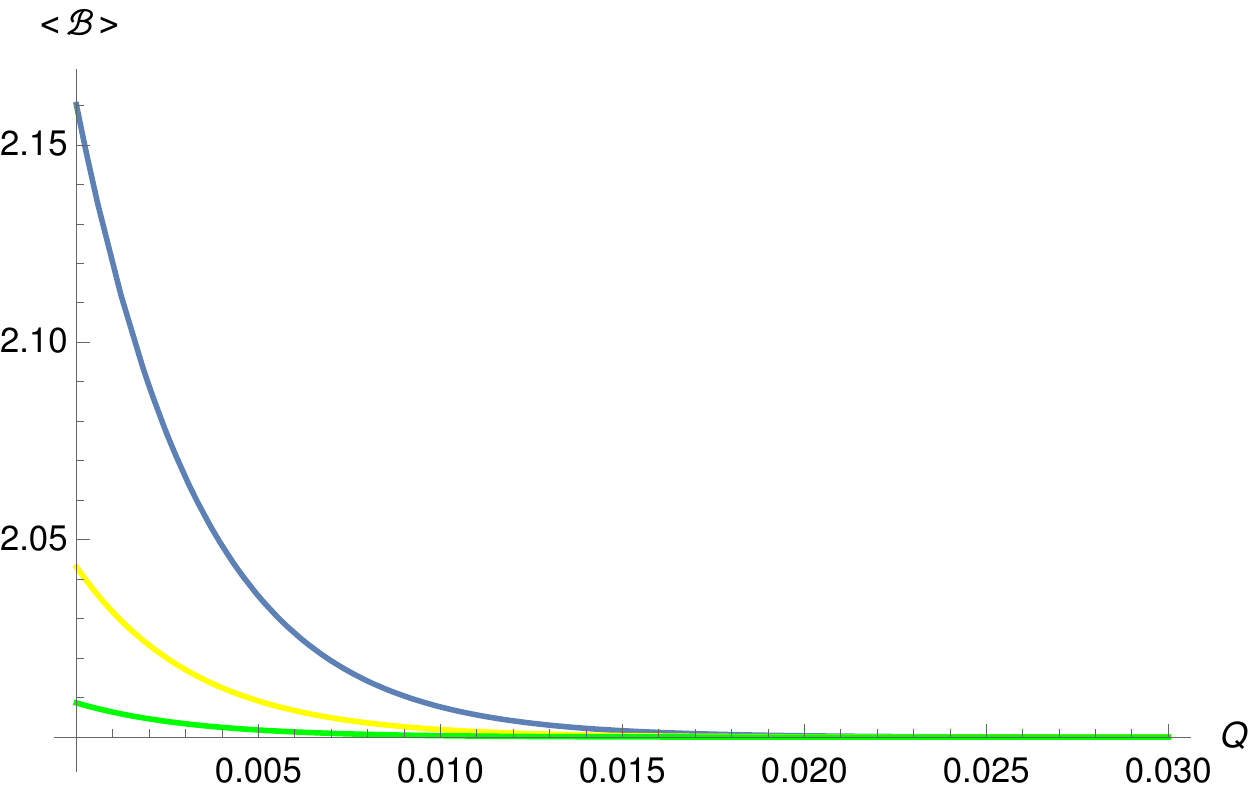}\hspace{1.0 cm}
\caption{We have plotted Bell's violation $\langle\mathcal{B}\rangle_{\rm max}$ \ref{bvpm}, corresponding to the vacuum state, with respect to the parameter   $Q=2enB/k_z^2$. We have plotted only the case $L^2\gg M^2$, for the other case ($M^2\gg L^2$) does not show any significant violation or numerical variation.   We have taken $L=100$ and different curves correspond to different $M$ values (blue $M=10$, yellow  $M=12$ and green  $M=16$). }
\label{fig:bellvofvacuum}
\end{figure}

Using the above constructions, and the squeezed state expansion \ref{vaccum1}  and also the operations \ref{spin} defined on the out states,  the desired expectation value is given by,
\begin{equation}
\label{B2}
^{\rm in}\langle 0^{(1)}_{k}|\mathcal{B}|0_{k}^{(1)}\rangle^{\rm in} = [E(\theta_{1},\theta_{2})+E(\theta_{1},\theta'_{2})+E(\theta'_{1},\theta_{2})-E(\theta'_{1},\theta'_{2})]
\end{equation}
where, $\mathit{O}_i$ and $\mathit{O}'_i$ are assumed to operate respectively on the $k$ and $-k$ sectors of the out states in \ref{vaccum1}, and 
$$E(\theta_{1},\theta_{2})
=
\cos\theta_{1} \cos\theta_{2} + 2|\alpha_{1}\beta_{1}| \sin\theta_{1} \sin\theta_{2}$$
Choosing now $\theta_{1}=0,\,\theta'_{1}=\pi/2$ and $\theta_{2}=-\theta'_{2}$, we have from \ref{B2},
\begin{equation}
^{\rm in}\langle 0_{k}^{(1)}|\mathcal{B}|0_{k}^{(1)}\rangle^{\rm in} = 2(\cos\theta_{2}+ 2|\alpha_{1}\beta_{1}|\sin\theta_{2})
\label{bvpm}
\end{equation}
The above expression maximises at $\theta_{2} =\tan^{-1}(2|\alpha_{1}\beta_{1}|)$, so that the above expectation value becomes
$$\langle\mathcal{B}\rangle_{\rm max} = 2\left(1+ 4|\alpha_1 \beta_1|^2 \right)^{1/2}$$
Thus $\langle\mathcal{B}\rangle_{\rm max}\geq2$, and hence there is  Bell violation for $|\beta_1|>0$.
We have plotted $\langle\mathcal{B}\rangle_{\rm max}$ in~\ref{fig:bellvofvacuum}  with respect to the parameter $Q=2enB/k_z^2$ as earlier.  We have considered only the case of strong electric field, $L^2\gg M^2$, for the other case does not show any significant violation nor numerical variation. As of the vacuum entanglement entropy, \ref{fig:EEforVacuum1}, the Bell violation decreases monotonically with the increasing magnetic field (for a given mode) and reaches the value two. Once again this happens due to the suppression of the particle creation by the magnetic field.

Note that the vacuum state is pure. Instead of vacuum, if we consider a pure but maximally entangled state, make its squeezed state expansion, and then trace out some parts of it in order to construct a bipartite subsystem, the resulting density matrix turns out to be mixed. The above construction is valid for pure ensembles only and one requires a different formalism to deal with mixed ensembles, e.g.~\cite{bell_3, nper:2011}. We wish to study two such cases below, in order to demonstrate their qualitative differences with the vacuum case. 

\subsection{Bell violation for maximally entangled initial states} \label{bvme}

We wish to consider maximally entangled initial states (corresponding to two fermionic fields) in the following.   For computational simplicity, we assume that both the fields have the same rest mass, and we consider modes in which their momenta along the $z$-direction  and the Landau levels are the same. 

The density matrix corresponding to the initial state can be expanded into the out states via \ref{vaccum1} and then any two degrees of freedom is traced out in order to construct a bipartite system. The resulting reduced density matrix turns out to be mixed.  For such a system, the Bell violation measure is defined as~\cite{bell_3, nper:2011},  
\begin{equation}
    \langle \mathcal{B}_{\rm max}\rangle\;=\; 2\sqrt{\lambda_{1}+\lambda_{2}},
    \label{bv}
\end{equation}
where $\lambda_{1}$ and $\lambda_{2}$ are the two largest eigenvalues of the $3\times 3$ matrix $U=(T_{\rho})^{\rm T}T_{\rho}$, with  $T \equiv {\rm Tr}[\rho \sigma_{i}\otimes \sigma_{j}]$, where $\rho$ is the aforementioned mixed density matrix. $T$ is called the correlation matrix for the generalised Bloch decomposition of $\rho$. Since the reduced density matrix represents a bipartite system, the violation of the Bell inequality as earlier will correspond to $ \langle \mathcal{B}_{\rm max}\rangle>2$ in \ref{bv}. \\

We begin by considering the initial state,
\begin{eqnarray}
\label{Bell-1}
|\psi\rangle&=&\frac{|0_p 0_{-p}0_k0_{-k}\rangle^{\text{in}}+|1_p0_{-p}0_{k}1_{-k}\rangle^{\text{in}}}{\sqrt{2}}
\end{eqnarray}
In the four entries of a ket above, the first pair of states corresponds to one fermionic field, whereas last pair corresponds to another. The $\pm$-sign in front of the momenta stands respectively for the particle and anti-particle degrees of freedom.  

Recall that we are assuming the created particles have the same rest mass, and we are working with modes for which  the Landau levels and the  $k_z$ values for both the fields are coincident.   Using then \ref{in_out}, \ref{vaccum1} we re-express \ref{Bell-1} in the out basis as
\begin{eqnarray}
|\psi\rangle=\frac{(\alpha_{1}|0_p0_{-p}\rangle^{\rm out}+\beta_{1}|1_p1_{-p}\rangle^{\rm out})(\alpha_{1}|0_k0_{-k}\rangle^{\rm out}+\beta_{1}|1_k1_{-k}\rangle^{\rm out})+|0_p1_{-p}\rangle^{\rm out}|1_{k}0_{-k}\rangle^{\rm out}}{\sqrt{2}}
\end{eqnarray}

We shall focus below only on the correlations between the particle-particle and the particle-anti-particle sectors corresponding to the density matrix of the above state. Accordingly, tracing out first the anti-particle-anti-particle degrees  of freedom of the density matrix $\rho^{(0)}\;=\;|\psi\rangle \langle \psi|$, we construct the reduced density matrix for the particle-particle sector,  
\begin{eqnarray}
\rho_{k,p}^{0}=\;\text{Tr}_{-k,-p}(\rho^{(0)})\;=\;\frac{1}{2}\left(\begin{array}{cccc}
|\alpha_{1}|^{4}&0&0&0\\
0&|\alpha_1\beta_1|^2&(\alpha_1\beta_1)^* &0\\
0&\alpha_1\beta_1&|\alpha_1\beta_1|^2&0\\
0&0&0&|\beta_{1}|^{4}\\
\end{array}\right)
\label{pp}
\end{eqnarray}
Likewise we can obtain the reduced density matrix for the particle-anti-particle sector,
\begin{eqnarray}
\rho_{p,-k}^{0}\;=\;\text{Tr}_{-p,k}(\rho^{(0)})=\frac{1}{2}\left( {\begin{array}{cccc}
|\alpha_{1}|^{4}&0&0&(\alpha_{1}^{*})^2\\
0&|\alpha_1\beta_1|^2&0&0\\
0&0&|\alpha_1 \beta_1|^2&0\\
\alpha_{1}^2&0&0&|\beta_{1}|^{4}+1\\
\end{array}}\right)
\label{ap}
\end{eqnarray}

The correlation matrices corresponding to \ref{pp} and  \ref{ap} are respectively given by,
\begin{eqnarray}
\label{correlation1}
T(\rho_{k,p}^{0})=\left( {\begin{array}{cccc}
	{\rm Re}(\alpha_1 \beta_1)&0&0\\
	0&-{\rm Re}(\alpha_1 \beta_1)&0\\
	0&0&\frac{1}{2}\left(|\alpha_{1}|^{4}-2|\alpha_1|^2|\beta_1|^2-1+|\beta_{1}|^4\right)\\
	\end{array}}\right)
\end{eqnarray}
and
\begin{eqnarray}
\label{correlation2}
T(\rho_{p,-k}^{0})=\left( {\begin{array}{cccc}
	{\rm Re}(\alpha_1^2)&0&0\\
	0&{\rm Re}(\alpha_1^2)&0\\
	0&0&\frac{1}{2}\left(|\alpha_{1}|^{4}-2|\alpha_1|^2|\beta_1|^2 +1+|\beta_1|^4\right)\\
	\end{array}}\right)
\end{eqnarray}
Using \ref{correlation1} and \ref{correlation2}, we compute the matrices $U(\rho_{k,p}^{0})=(T(\rho_{k,p}^{0}))^{\rm T}T(\rho_{k,p}^{0})$, and  $U(\rho_{p,-k}^{0})=(T(\rho_{p,-k}^{0})^{\rm T}T(\rho_{p,-k}^{0})$.   \ref{bv} yields then after a little algebra 
\begin{equation}
\label{Bell0kp}
     \langle \mathcal{B}^{0}_{kp}\rangle_{\rm max} = 2\sqrt{2}\,{\rm Re}(\alpha_1 \beta_1)
\end{equation}
and
\begin{equation}
\label{Bell0p-k}
     \langle \mathcal{B}^{0}_{p-k}\rangle_{\rm max} = 2\sqrt{({\rm Re}(\alpha_1^2))^2+\left(1-2|\alpha_1\beta_1|^2\right)^2}
\end{equation}

\begin{figure}[h]
\centering
\includegraphics[scale=0.45]{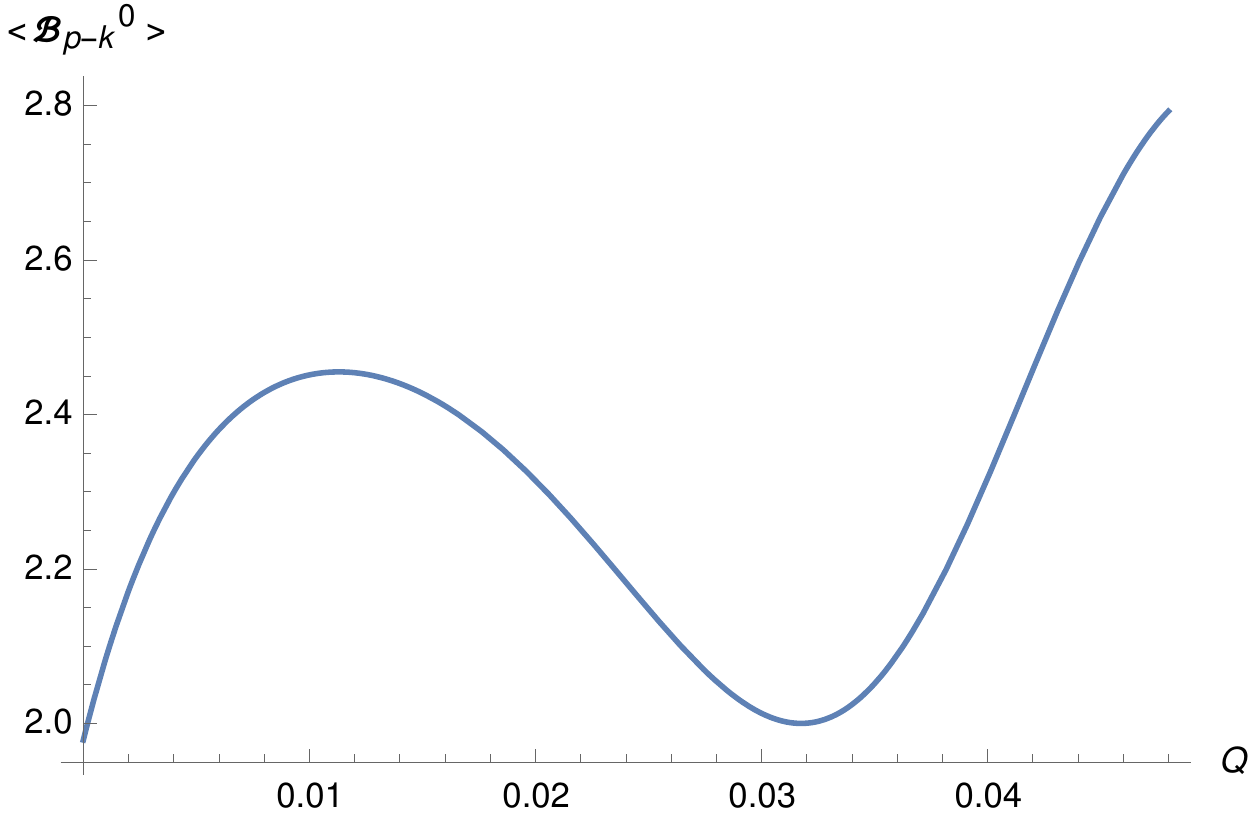}\hspace{1.0cm}
\includegraphics[scale=0.45]{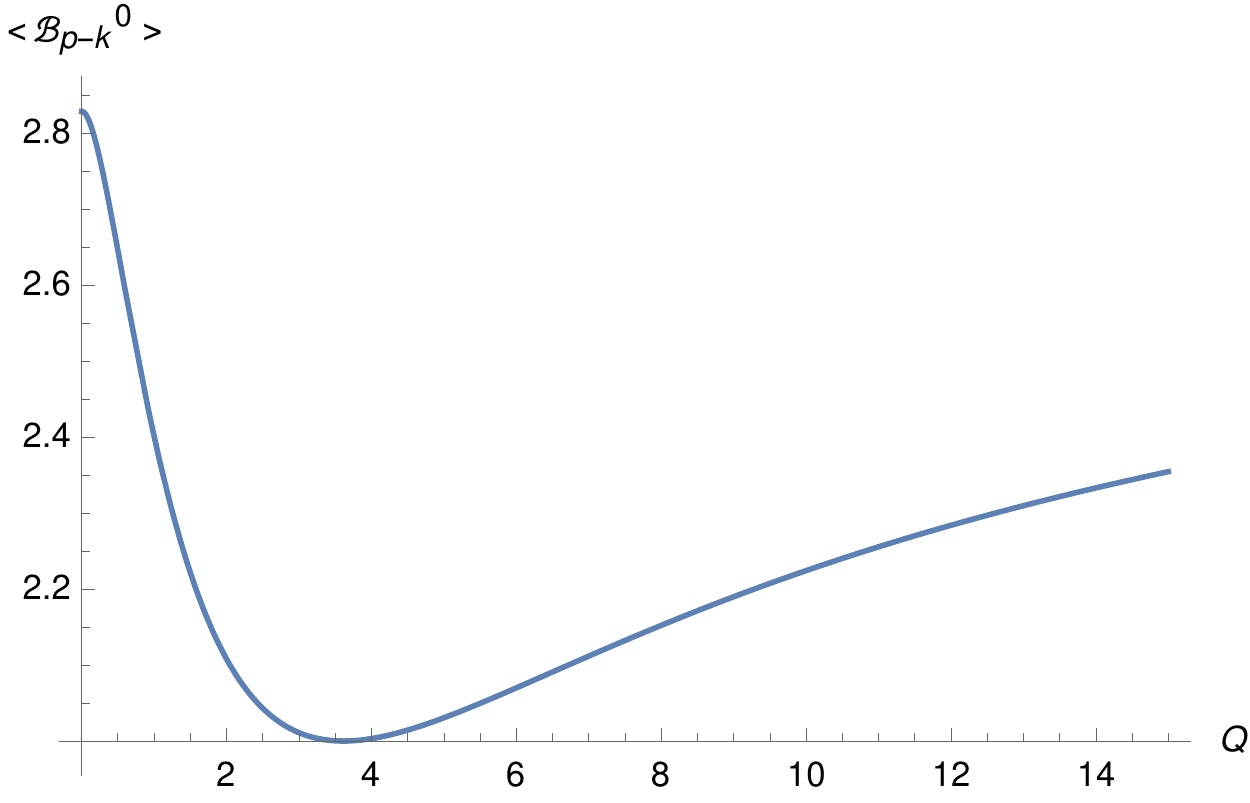}
\caption{Bell violation for the particle-anti-particle sector $\rho_{p,-k}^{0}$, \ref{ap}, corresponding to the initial state in \ref{Bell-1}.  We have plotted  \ref{Bell0p-k} with respect to the parameter $Q=2neB/k_z^2$.  The left plot corresponds to $L^2\gg M^2$  ($L=100$ and $M=10$), whereas the right one corresponds to  $M^2\gg L^2$ ($M=5$ and $L=1$). $\langle \mathcal{B}^{0}_{p-k}\rangle_{\rm max}>2$ corresponds to the Bell violation. 
}
\label{fig:rhop-k0}
\end{figure}

 We have plotted $\langle \mathcal{B}^{0}_{p-k}\rangle_{\rm max}$ in~\ref{fig:rhop-k0} with respect to the parameter $Q=2neB/k_z^2$ as earlier, depicting the Bell violation ($\langle \mathcal{B}^{0}_{p-k}\rangle_{\rm max}>2$) for both strong and weak electric fields.  $\langle \mathcal{B}^{0}_{kp}\rangle_{\rm max}$ on the other hand, does not show any such violation.

\bigskip

We next consider another maximally entangled state given by,
\begin{eqnarray}
\label{Bell-2}
|\chi\rangle=\frac{|1_p 0_{-p}0_k0_{-k}\rangle^{\text{in}}+|0_p0_{-p}1_{k}0_{-k}\rangle^{\text{in}}}{\sqrt{2}}
\end{eqnarray}

Following similar steps as described above, by partially tracing out  the original density matrix $\rho^{(1)}=|\chi\rangle \langle \chi |$, we have the mixed bipartite density matrices respectively for the particle-particle and the particle-anti-particle sectors,
\begin{equation}
\rho_{k,p}^{1}\;=\;\text{Tr}_{-k,-p}(\rho^{(1)}) \qquad {\rm and} \qquad 
\rho_{p,-k}^{1}\;=\;\text{Tr}_{-p,k}(\rho^{(1)}),
\label{z}
\end{equation}
which respectively yield the Bell violations,
\begin{equation}
     \langle \mathcal{B}^{1}_{kp}\rangle_{\rm max} = 2\sqrt{2}|\alpha_1|^2 \qquad 
{\rm and} \qquad 
     \langle \mathcal{B}^{1}_{p-k}\rangle_{\rm max} = 2\sqrt{2}\,{\rm Re}(\beta_1\alpha^{*}_1)
     \label{Bell1p-k}
\end{equation}

\begin{figure}[h]
	\begin{center}
		\includegraphics[scale=.55]{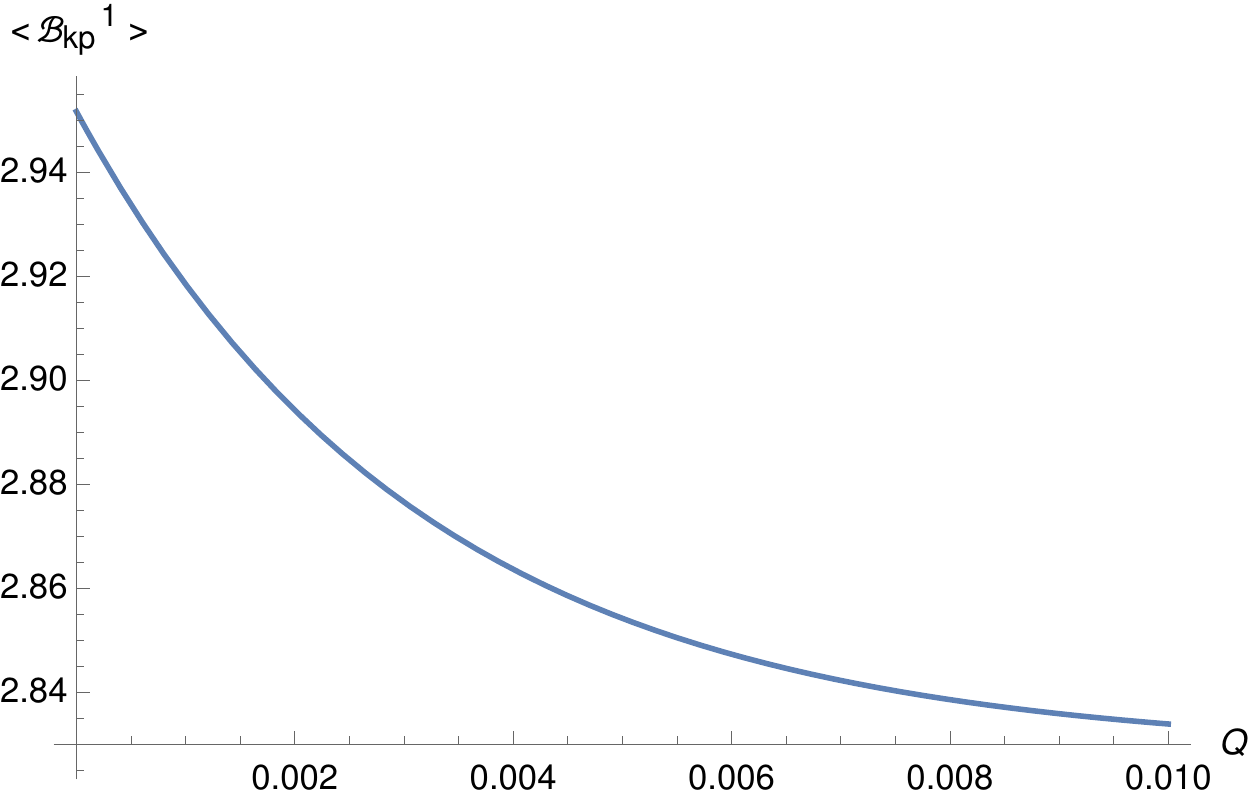}\hspace{1.0 cm}
		\caption{
		Bell violation for $\rho_{k,p}^{1}$, \ref{z}, corresponding to the initial state in \ref{Bell-2}. We have plotted $\langle \mathcal{B}^{1}_{kp}\rangle$, \ref{Bell1p-k},  with respect to the parameter $Q=2neB/k_z^2$, for $L^2\gg M^2$ ($L=100$ and $M=10$). The other case, $M^2 \gg L^2$ does show Bell violation but there is no significant numerical variation. Note that in contrast to \ref{fig:rhop-k0} the behaviour is monotonic here and qualitatively it rather resembles with the vacuum case, \ref{fig:bellvofvacuum}.
		}
		\label{fig:rhokp1}
		\end{center}
\end{figure}

We have plotted $ \langle \mathcal{B}^{1}_{kp}\rangle_{\rm max}$ in \ref{fig:rhokp1} with respect to the parameter $Q$ for strong electric field, $L^2 \gg M^2$. For $M^2\gg L^2$, we also have violation, however it does not show any significant numerical variation. 
 On the other hand, we find no violation for the particle-anti-particle sector, $\langle \mathcal{B}^{1}_{p-k}\rangle_{\rm max}$.

Before we conclude, we wish to further extend the above results for the so called fermionic $\alpha$-vacua.

\section{The case of the fermionic $\alpha$-vacua}\label{alph}
The fermionic $\alpha$-vacua, like the scalar field~\cite{Allen, alpha-vaccuum1}, correspond to a Bogoliubov transformation characterised by a parameter $\alpha$ in the in mode field quantisation.  Although such vacua may not be very useful to do perturbation theory, e.g.~\cite{Mottola:1984ar, Einhorn}, it still attracts attention chiefly from the perspective of the so called trans-Planckian censorship conjecture,~e.g~\cite{Bedroya}. 

In order to construct such vacua, from  \ref{field}, we define a new set of annihilation and creation operators \cite{alpha-vacua},
\begin{eqnarray}
    \label{alphaop1}
    c^{\alpha}({\vec{k}_{\slashed{y}},s,n})
    &=&\cos\alpha\, a_{\rm in}(\vec{k}_{\slashed{y}},s,n)-\sin \alpha\, b^{\dagger}_{\rm in}(\vec{k}_{\slashed{y}},s,n) \nonumber\\
   d^{\alpha}({\vec{k}_{\slashed{y}},s,n})
&=&\cos\alpha\;b_{\rm in}(\vec{k}_{\slashed{y}},s,n)+\sin\alpha\;a^{\dagger}_{\rm in}(\vec{k}_{\slashed{y}},s,n)  
\end{eqnarray}
where the parameter $\alpha$ is real and $0 \leq \alpha \leq \pi/2$. The above relations indicate that we need to define a new, one parameter family of vacuum state $|0\rangle_{\alpha}$, so that
$$ c^{\alpha}|0\rangle_{\alpha} =0=d^{\alpha}|0\rangle_{\alpha}$$
An $\alpha$-vacuum state is related to the original in-vacuum state via a squeezed state expansion. Note that \ref{alphaop1} does not mix the sectors $s=1$ and $s=2$. Thus as of the previous analysis, we work only with the $s=1$ sector and write for the normalised $\alpha$-vacuum state,
\begin{equation}
\label{alphavacuum}
    |0_k\rangle^{(1)}_{\alpha} = \cos{\alpha}|0_k^{(1)} 0_{-k}^{(1)} \rangle^{\rm in} + \sin{\alpha}|1_k^{(1)} 1_{-k}^{(1)} \rangle^{\rm in} 
\end{equation}
Using now \ref{vaccum1} into the above equation, we re-express $|0\rangle^{(1)}_{\alpha} $ in terms of the out states,
\begin{equation}
    \label{alpha_out}
    |0_{k}\rangle^{(1)}_{\alpha} = \alpha' |0_k^{(1)} 0_{-k}^{(1)}\rangle^{\rm out} + \beta' |1_k^{(1)}1_{-k}^{(1)}\rangle^{\rm out}
\end{equation}
where 
\begin{eqnarray}
\alpha'=\frac{\alpha_1\cos{\alpha}+\beta_1 \sin{\alpha}}{\sqrt{1+2(\alpha_1 \beta_1^* + \beta_1 \alpha_1^*)\cos{\alpha}\sin{\alpha}}}, \qquad \beta'=\frac{\alpha_1\sin{\alpha}+\beta_1 \cos{\alpha}}{\sqrt{1+2(\alpha_1 \beta_1^* + \beta_1 \alpha_1^*)\cos{\alpha}\sin{\alpha}}}
\label{b'}
\end{eqnarray}
 are the effective Bogoliubov coefficients. Note the formal similarity between \ref{alpha_out} and \ref{vaccum1}. Setting $\alpha=0$ in the first reproduces the second.\\

\noindent
The above mentioned formal similarity thus ensures that the expressions for either the vacuum entanglement entropy or the Bell violation for the $\alpha$-states can be obtained from our earlier results, \ref{entropy}, \ref{bvpm}, \ref{Bell0kp}, \ref{Bell0p-k}, \ref{Bell1p-k}, by simply making the replacements,
$$\alpha_1 \to \alpha'_1, \qquad {\rm and} \qquad \beta_1 \to \beta'_1$$
Some aspects of entanglement for scalar $\alpha$-vacua can be seen in, e.g.~\cite{EE for alpha1, EE for alpha, BI for mixed state, Choudhury:2017qyl} (also references therein). See also \cite{SSS} for discussion on the natural emergence of $\alpha$-like vacua for fermions in the hyperbolic de Sitter spacetime. 

\begin{figure}[h]
	\begin{center}
		\includegraphics[scale=.6]{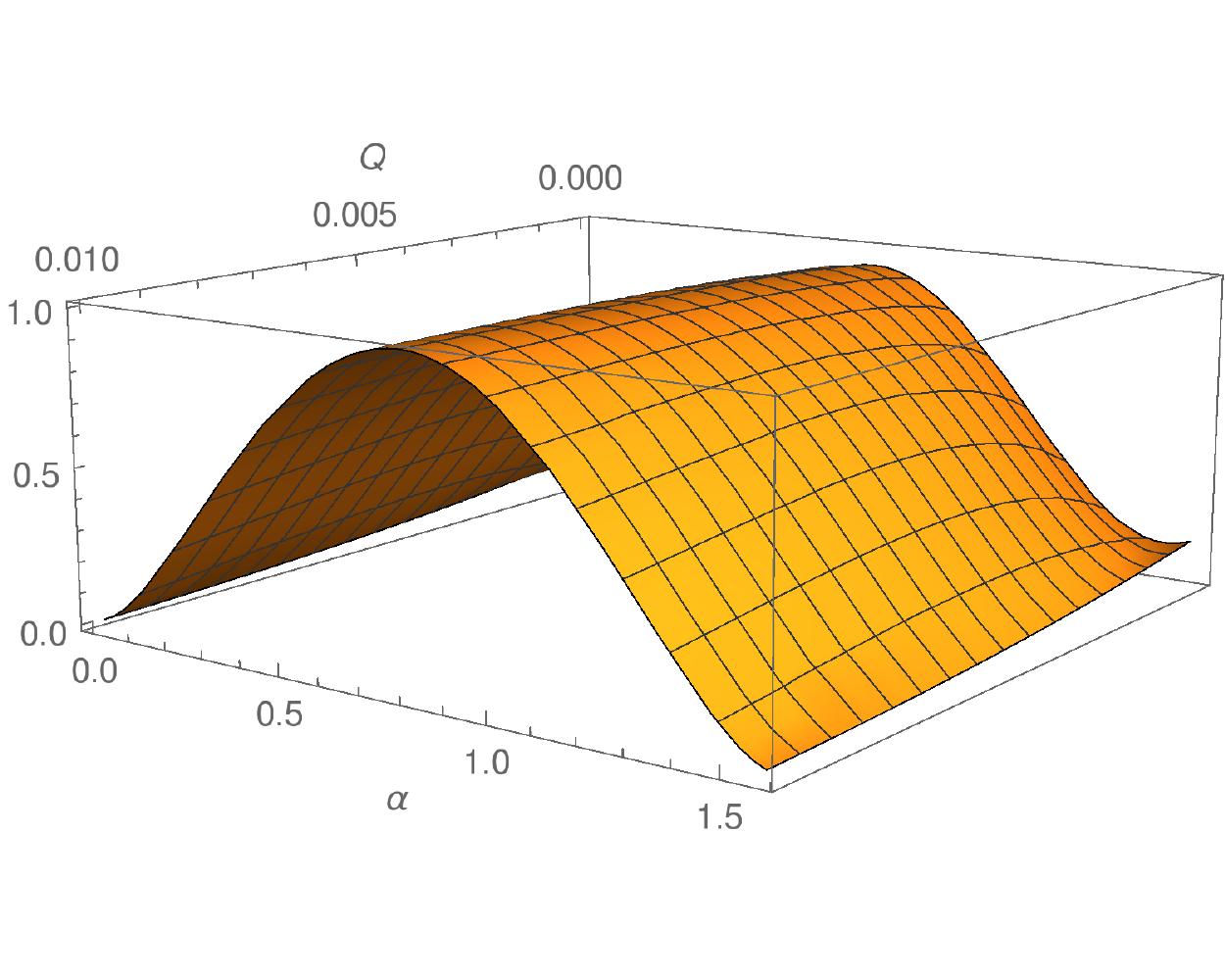}\hspace{1.0 cm}
			\includegraphics[scale=.55]{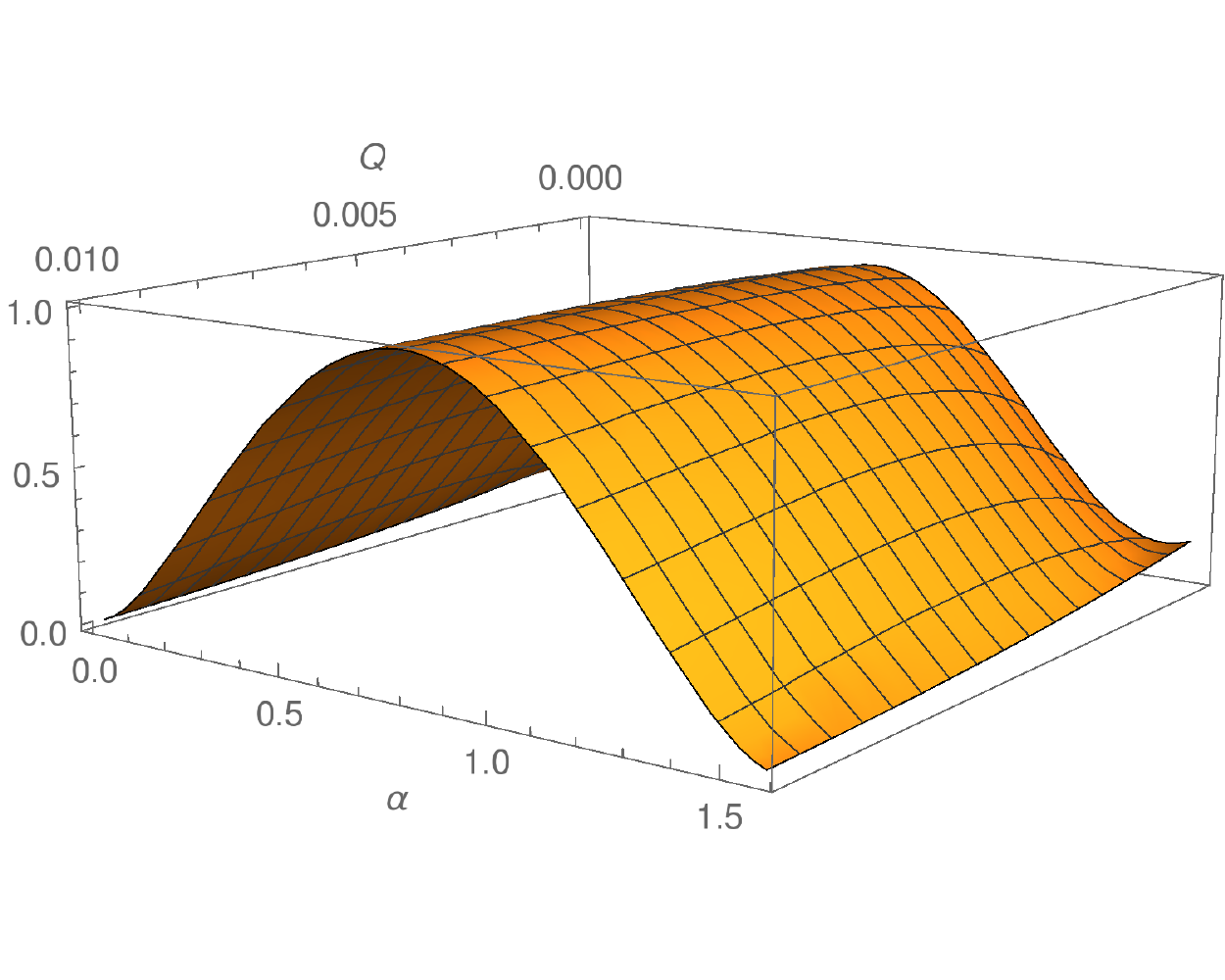}
		\caption{
		Entanglement entropy for the femionic  $\alpha$-vacua, \ref{alphavacuum}. We have plotted   \ref{entropy'} vs. the parameters $Q=2enB/k_z^2$ and $\alpha$. The left one corresponds to $L^2\gg M^2$ ($L=100$ and $M=10$), whereas the right one corresponds to $M^2\gg L^2$ ($M=5$ and $L=1$). See main text for discussions.
		 	}
		\label{fig:alphaentropy}
		\end{center}
\end{figure}

For the fermionic case the vacuum entanglement entropy, \ref{entropy}, modifies to the $\alpha$-vacua as,
\begin{equation}
    \label{entropy'}
	S_k^{\alpha} =
		-   \left[ \ln	(1-|\beta'_1|^{2})    +|\beta'_1|^2  \ln \frac{|\beta'_1|^{2}}{1-|\beta'_1|^2}\right]
\end{equation}
which is plotted in \ref{fig:alphaentropy} with respect to the parameters $Q=2enB/k_z^2$ and $\alpha$. We see that $S_k^{\alpha}$ first increases with increase in the parameter $\alpha$ and has its maximum at $\alpha = \pi/4$, after which it decreases and becomes vanishing as $\alpha \to \pi/2$.
\begin{figure}[h]
	\begin{center}
		\includegraphics[scale=.55]{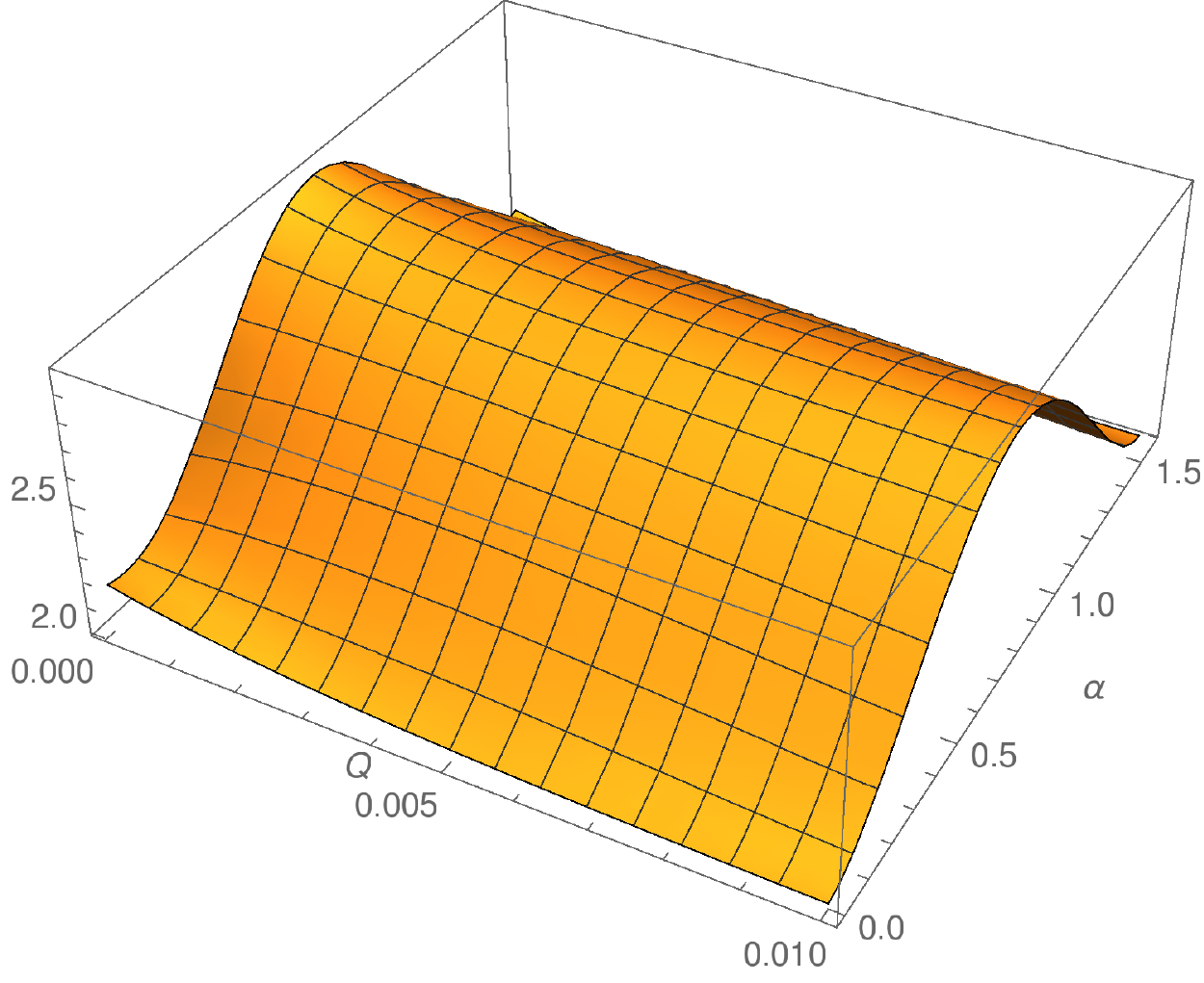}\hspace{1.0cm}
		\includegraphics[scale=.55]{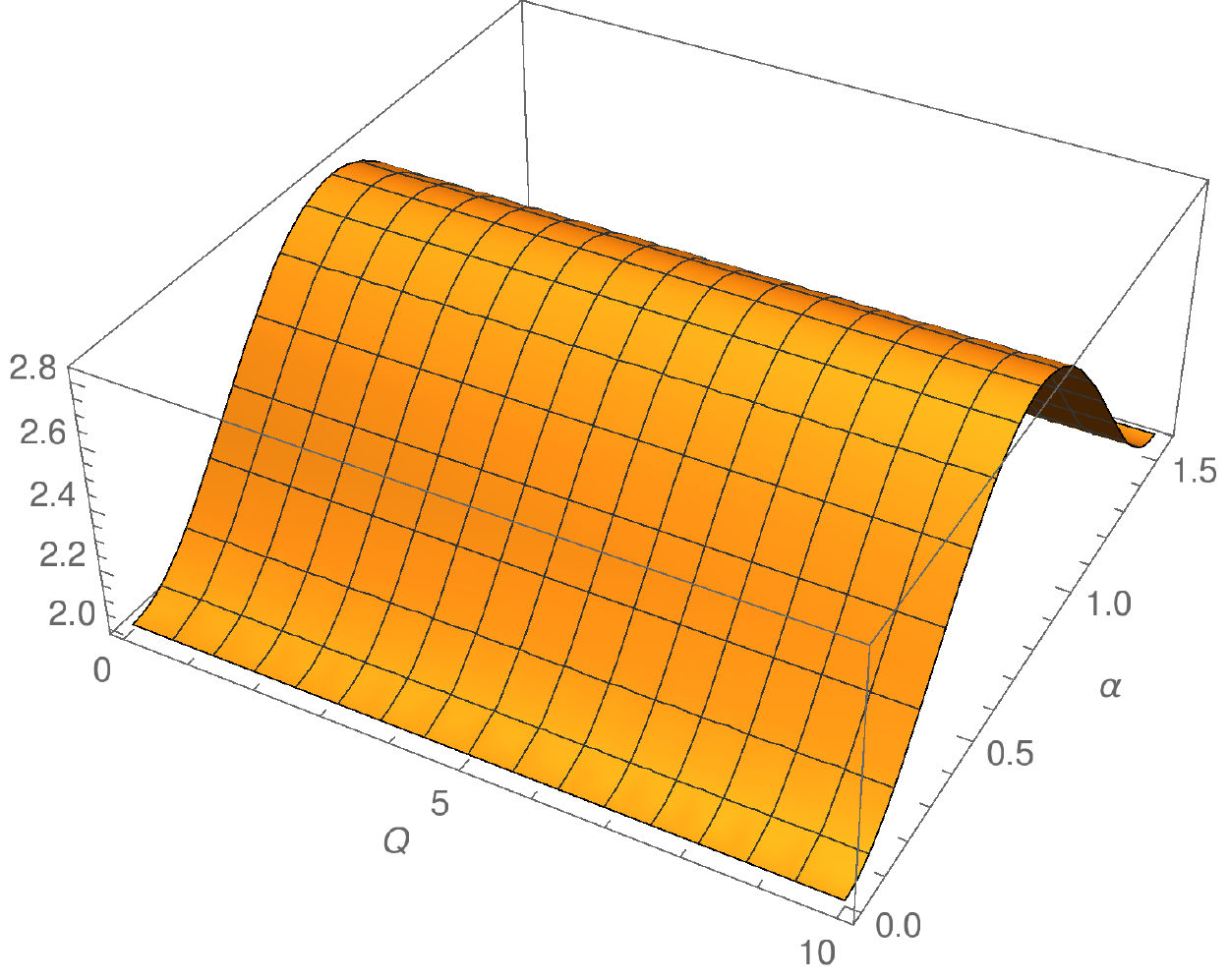}
		\caption{	Bell violation for the fermionic $\alpha$-vacua. As we have discussed in the main text,  we have plotted   \ref{bvpm} after replacing $\alpha_1$ and $\beta_1$, respectively by $\alpha'_1$ and $\beta'_1$ given by \ref{b'}. The left one corresponds to $L^2\gg M^2$ ($L=100$ and $M=10$), whereas the right one corresponds to $M^2\gg L^2$ ($M=5$ and $L=1$). $\langle \mathcal{B}\rangle_{\alpha,\,\rm max}>2$ corresponds to the Bell violation.
			}
		\label{fig:BValphavaccuum}
		\end{center}
\end{figure}
Likewise the Bell violation for the vacuum state, \ref{bvpm}, can be extended to the $\alpha$-vacua and is plotted in \ref{fig:BValphavaccuum}. Like the vacuum entanglement entropy, the vacuum Bell violation also reaches maximum at $\alpha = \pi/4$ and becomes vanishing as $\alpha \to \pi/2$. 

The vanishing of both vacuum entanglement entropy and Bell violation as $\alpha\to\pi/2$ can be understood as follows. In this limit, only the excited state part of \ref{alphavacuum} survives.  \ref{vaccum1} then implies that the corresponding out-basis expansion of this state is not only pure, but also separable. Thus in this limit no entanglement survives.\\

Let us now come to the case of the maximally entangled states. The states of \ref{Bell-1}, \ref{Bell-2} respectively modify as,
\begin{eqnarray}
\label{Bell-1a}
|\psi\rangle_{\alpha}=\frac{|0_p 0_{-p}0_k0_{-k}\rangle_{\alpha}^{\text{in}}+|1_p0_{-p}0_{k}1_{-k}\rangle_{\alpha}^{\text{in}}}{\sqrt{2}}
\end{eqnarray}
and,
\begin{eqnarray}
\label{Bell-2a}
|\chi\rangle_{\alpha}=\frac{|1_p 0_{-p}0_k0_{-k}\rangle_{\alpha}^{\text{in}}+|0_p0_{-p}1_{k}0_{-k}\rangle_{\alpha}^{\text{in}}}{\sqrt{2}}
\end{eqnarray}

\begin{figure}[h]
	\begin{center}
		\includegraphics[scale=.55]{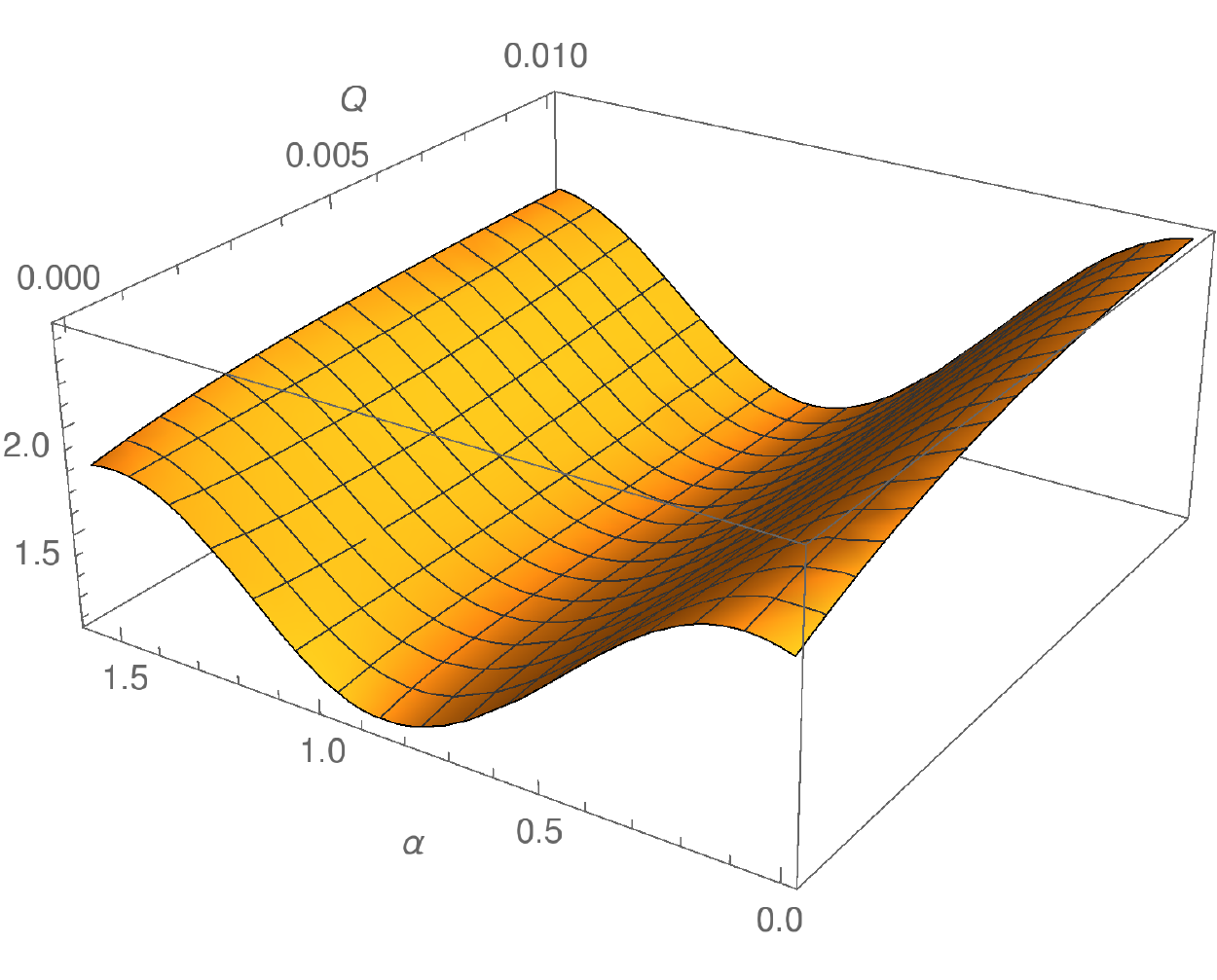}\hspace{1.0cm}
		\includegraphics[scale=.55]{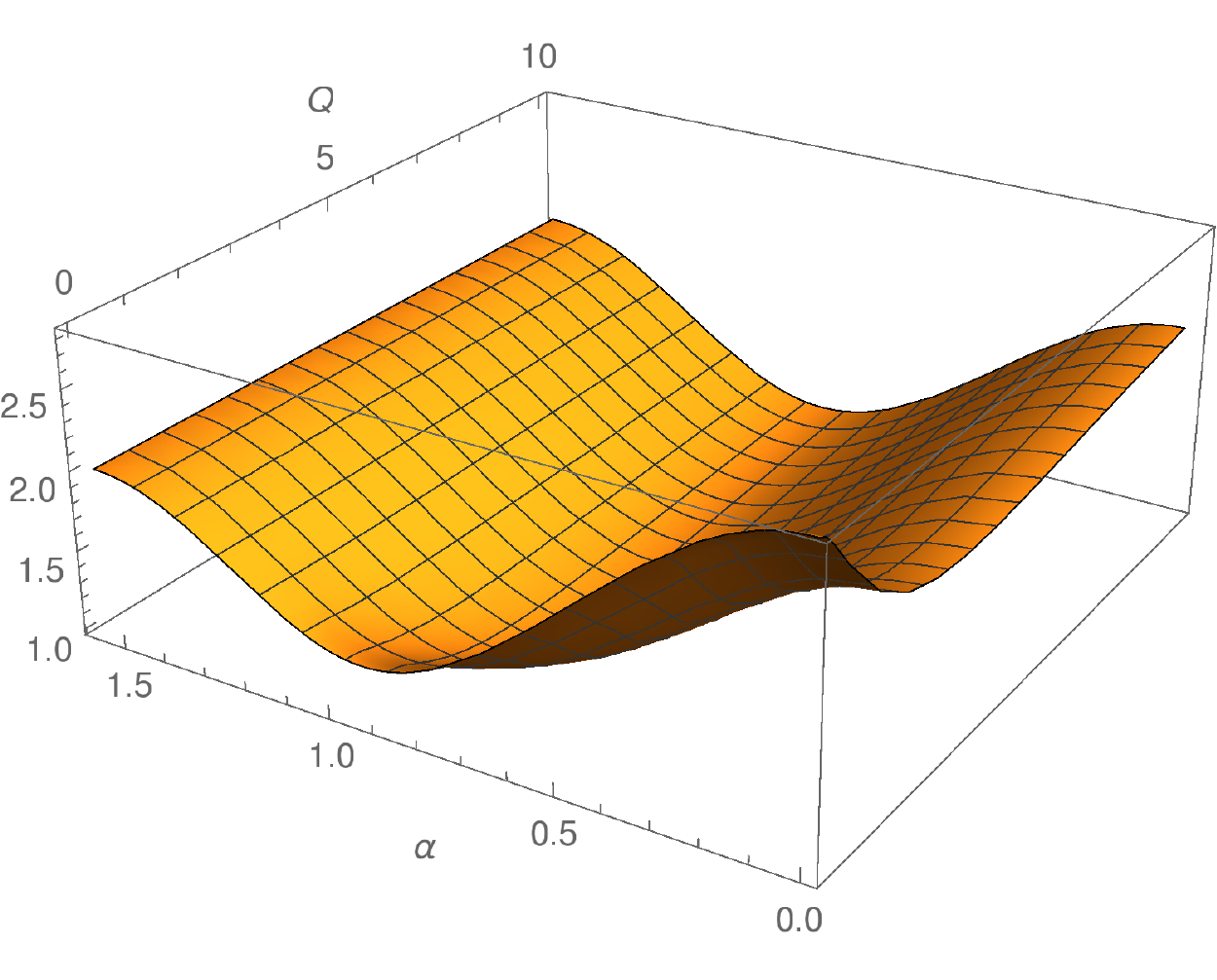}
		\caption{Bell violation for the particle-anti-particle sector corresponding to the initial $\alpha$-state in \ref{Bell-1a}. As we have discussed in the main text, we have basically plotted \ref{Bell0p-k} after replacing  $\alpha_1$, $\beta_1$ respectively by $\alpha'_1$ and  $ \beta'_1$ (\ref{b'}).  The left plot corresponds to $L^2\gg M^2$  ($L=100$ and $M=10$), whereas the right one corresponds to  $M^2\gg L^2$ ($M=5$ and $L=1$). $\langle \mathcal{B}\rangle_{\alpha,\,{\rm max}}>2$ corresponds to the Bell violation. The particle-particle sector corresponding to this initial state does not show any Bell violation, like the  $\alpha=0$ case discussed in \ref{bvme}.
			}
		\label{fig:BValpharhop-k0}
		\end{center}
\end{figure}

\begin{figure}[h]
	\begin{center}
		\includegraphics[scale=.55]{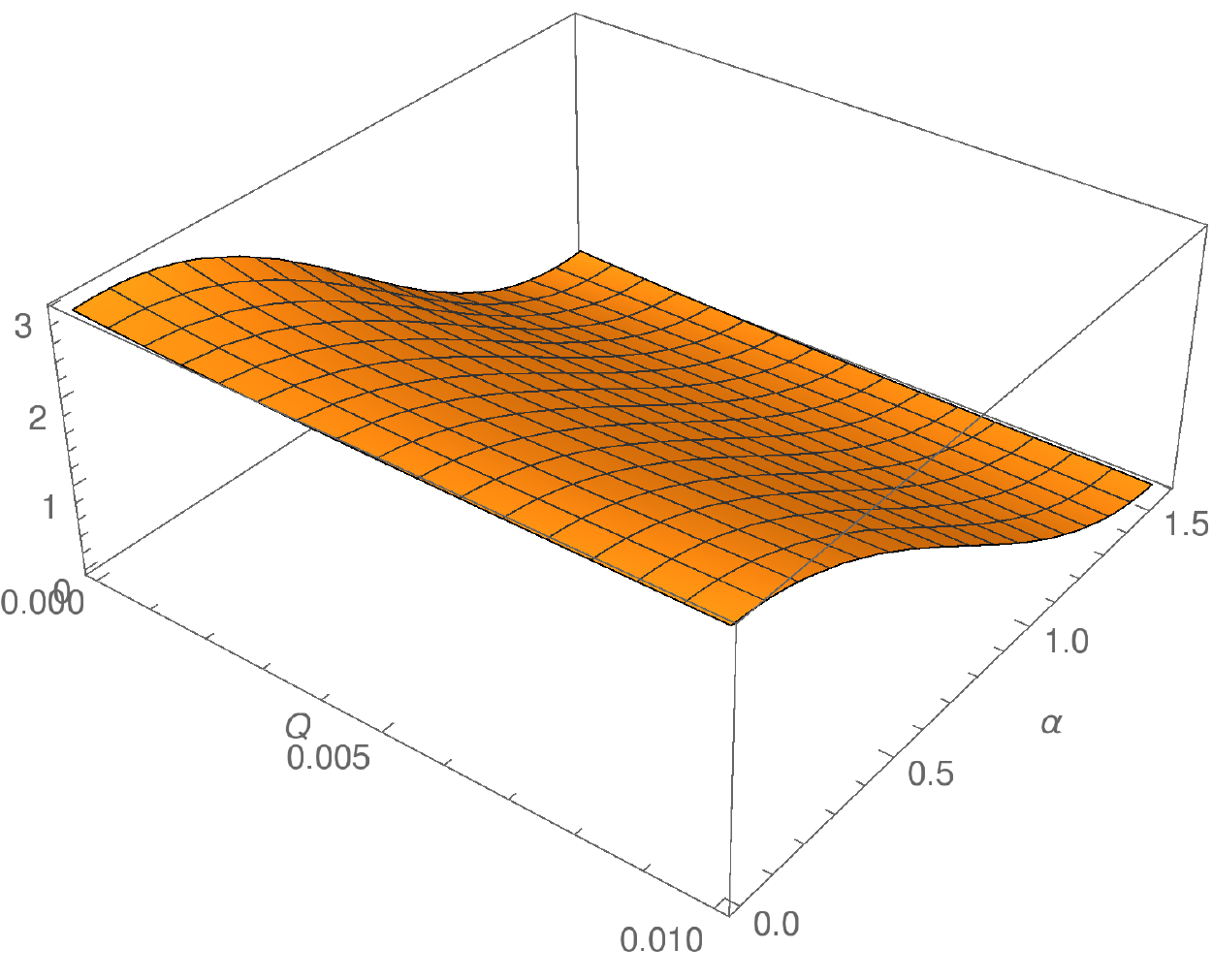}\hspace{1.0cm}
		\includegraphics[scale=.55]{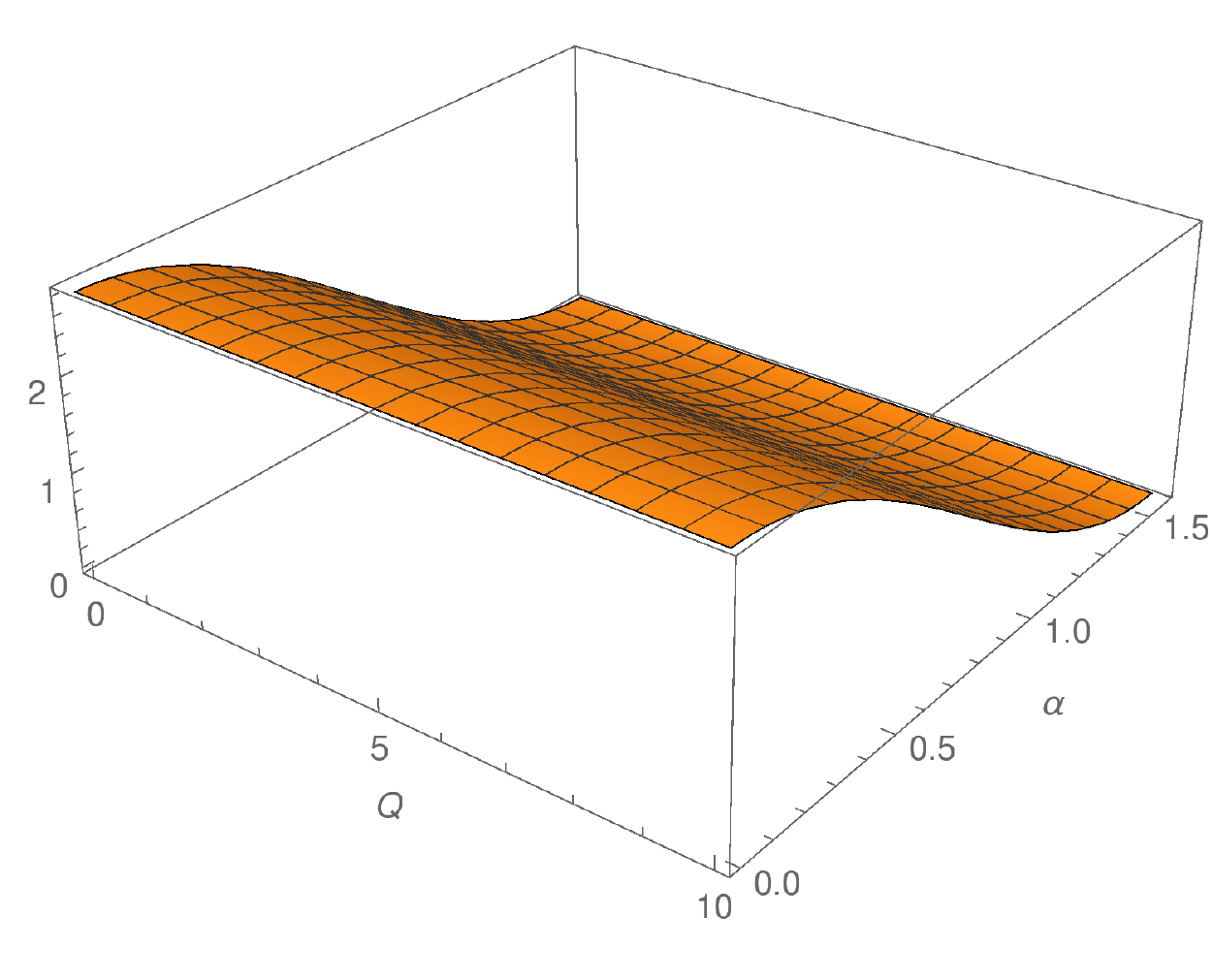}
		\caption{	Bell violation for the particle-particle sector corresponding to the initial state in \ref{Bell-2a}. The left plot corresponds to $L^2\gg M^2$  ($L=100$ and $M=10$), whereas the right one corresponds to  $M^2\gg L^2$ ($M=5$ and $L=1$). As earlier, $\langle \mathcal{B}\rangle_{\alpha,{\rm max}}>2$ corresponds to the Bell violation. The particle-anti-particle sector corresponding to this initial state does not show any violation, like the $\alpha=0$ case discussed in \ref{bvme}.
			}
		\label{fig:BValpharhokp1}
		\end{center}
\end{figure}

Using \ref{alphavacuum}, and the method described in~\ref{bvme}, we can easily extend the results of the Bell violation we found earlier. As we mentioned earlier, this generalisation effectively corresponds to  just replacing  $\alpha_1$, $\beta_1$ respectively by $\alpha'_1$ and  $ \beta'_1$ (\ref{b'}) in appropriate places  (e.g. in \ref{Bell0kp}).   We have plotted these Bell violations in \ref{fig:BValpharhop-k0}, \ref{fig:BValpharhokp1}.
%

\section{Summary and outlook}
\label{sec:SD}
In this work we have discussed the fermionic Bell violation in the cosmological de Sitter spacetime, in the presence of primordial electromagnetic fields of constant strengths. We have found relevant in and out orthonormal Dirac  mode functions, the Bogoliubov coefficients and the resultant squeezed state relationship between the in and out states, in \ref{S2}.  Using these key results, we have computed  the vacuum entanglement entropy and the Bell violation (for both vacuum and two maximally entangled initial states) respectively in \ref{EE}, \ref{BV}. These results are extended further to the so called fermionic $\alpha$-vacua in \ref{alph}. We have focused on two qualitatively distinct cases here -- the `strong' electric filed and the `heavy' mass limits (with respect to the Hubble constant), c.f. \ref{case1}, \ref{case2}.  

As we have discussed in \ref{S1}, a background magnetic field alone cannot create vacuum instability, but in the presence of spacetime curvature and electric field, it can affect such instability or the rate of the particle pair creation. This is manifest from \ref{number_demsity}, which receives, as we have discussed,  no contribution from the magnetic field if the electric field strength is vanishing. Whereas if the magnetic field strength is very large compared to that of the electric field,  the particle creation rate also becomes independent of the electromagnetic fields.   Our chief aim in this paper was to investigate the role of the magnetic field strength on the Bell violation. We have seen that subject to the choices of the initial states, the behaviour of the Bell violation can be qualitatively different, e.g.~\ref{fig:bellvofvacuum} and \ref{fig:rhop-k0}. For the case of the $\alpha$-vacua on the other hand, we have also taken into account the variation of the parameter $\alpha$,~e.g.~\ref{fig:alphaentropy}.

The above analysis can be attempted to be extended in a few interesting scenarios. For example, instead of having only constant electromagnetic fields, can we also have fluctuating ones, like electromagnetic radiation? Can one also include the effect of gravitational radiation? Finally, it seems also interesting to perform similar analysis in the Rindler spacetime, for its relevance to the near horizon geometry of non-extremal black holes. Discussion of the Schwinger pair creation for a complex scalar field coupled to a constant background electric field in the Rindler spacetime can be seen in~\cite{Gabriel:1999yz}. Finally, as we have discussed in~\ref{S1}, it will be important to compute the breaking of scale invariance of the cosmological power spectra in the presence of primordial electromagnetic fields and also to compute the Bell violation by the photons (interacting with the entangled fermions)  coming from very distant sources, with the hope to constrain the strengths of those background fields. A  gauge invariant formulation of an effective action for the second problem seems  to be a non-trivial task. We hope to come back to this issue in future works. 

\bigskip

\section*{Acknowledgement}
MSA is fully supported by the ISIRD grant 9-252/2016/IITRPR/708. SB is partially supported by the ISIRD grant 9-289/2017/IITRPR/704. SC is
partially supported by the ISIRD grant 9-252/2016/IITRPR/708.

\appendix
\labelformat{section}{Appendix #1} 

\section{Explicit form of the mode functions and normalisations}\label{A}

The four orthonormal simultaneous eigenvectors $\omega_s$ of the operators $(M\gamma^0+L \gamma^0 \gamma^3)$ and $\gamma^1 \gamma^2$ appearing in \ref{dirac4} are given by,
\begin{eqnarray}
\label{eigenkets}
\omega_1&=&\frac{1}{P_1}\left( {\begin{array}{cccc}
   \frac{\sqrt{M^2+L^2}-L}{M}\\
0\\
1\\
   0\\
  \end{array} } \right), \qquad 
  \;\omega_2=\frac{1}{P_2}\left( {\begin{array}{cccc}
  0\\
 \frac{\sqrt{M^2+L^2}+L}{M}\\
0\\
   1\\
  \end{array} } \right), \nonumber\\
  \omega_3&=&\frac{1}{P_1}\left( {\begin{array}{cccc}
  0\\
 \frac{\sqrt{M^2+L^2}-L}{M}\\
0\\
   -1\\
  \end{array} } \right), \qquad
  \;\omega_4=\frac{1}{P_2}\left( {\begin{array}{cccc}
 \frac{\sqrt{M^2+L^2}+L}{M}\\
 0\\
-1\\
   0\\
  \end{array} } \right),
\end{eqnarray}
where $P_1$ and $P_2$ are normalisation constants. $\omega_3$ and $\omega_4$ are respectively related to $\omega_1$ and $\omega_2$ via the charge conjugation, $\omega_{3,4}= {\cal C}\, \omega^{*}_{1,2}$, where ${\cal C}=i\gamma^2$. The explicit representation of the gamma matrices we are using is given by,
\begin{eqnarray}
\gamma^0\;=\;\left(\begin{array}{ccc}
0&I\\
I&0\\
\end{array}\right), \qquad 
\gamma^i\;=\;\left(\begin{array}{ccc}
0&\sigma^i\\
-\sigma^i&0\\ 
\end{array}\right)\qquad (i=1,2,3)
\end{eqnarray}

We also note here the explicit forms of the positive frequency in and out modes appearing in \ref{mainmodes}, as follows,
\begin{equation}
\label{positive1in}
\begin{split}
U_{1,n}^{\rm in}=\frac{ \gamma^0}{N_1 a^{3/2}}\left\lbrace\left(i\partial_\eta-k_z\gamma^0 \gamma^3+aH\left(M \gamma^0 + L \gamma^0 \gamma^3 \right)\right)+\left(i\gamma^0 \gamma^2 \partial_y-\left(k_x+eBy\right)\gamma^0 \gamma^1\right)\right\rbrace \\ \times e^{-iHLz}e^{i\vec{k}\slashed{y}\cdot\vec{x}} W_{\kappa_1,i|\mu|}(z_1)h_{n}(\overline{y})\, \omega_1
  \end{split}
\end{equation}
\begin{equation}
\begin{split}
U_{2,n}^{\rm in}=\frac{\gamma^0}{N_2a^{3/2}} \left\lbrace\left(i\partial_\eta-k_z\gamma^0 \gamma^3+aH\left(M \gamma^0 + L \gamma^0 \gamma^3 \right)\right)+\left(i\gamma^0 \gamma^2 \partial_y-\left(k_x+eBy\right)\gamma^0 \gamma^1\right)\right\rbrace \\  \times e^{-iHLz}e^{i\vec{k}\slashed{y}\cdot\vec{x}} W_{\kappa_2,i|\mu|}(z_2)h_{n}(\overline{y})\, \omega_2
%
 \end{split}
\end{equation}
\begin{equation}
\label{positive1out}
\begin{split}
U_{1,n}^{\rm out}=\frac{\gamma^0}{M_1a^{3/2}} \left\lbrace\left(i\partial_\eta-k_z\gamma^0 \gamma^3+aH\left(M \gamma^0 + L \gamma^0 \gamma^3 \right)\right)+\left(i\gamma^0 \gamma^2 \partial_y-\left(k_x+eBy\right)\gamma^0 \gamma^1\right)\right\rbrace \\ \times e^{-iHLz}e^{i\vec{k}\slashed{y}\cdot\vec{x}} M_{\kappa_1,i|\mu|}(z_1)h_{n}(\overline{y})\,\omega_1
%
  \end{split}
\end{equation}
\begin{equation}
\begin{split}
U_{2,n}^{\rm out}=\frac{\gamma^0}{M_2 a^{3/2}} \left\lbrace\left(i\partial_\eta-k_z\gamma^0 \gamma^3+aH\left(M \gamma^0 + L \gamma^0 \gamma^3 \right)\right)+\left(i\gamma^0 \gamma^2 \partial_y-\left(k_x+eBy\right)\gamma^0 \gamma^1\right)\right\rbrace \\ \times  e^{-iHLz}e^{i\vec{k}\slashed{y}\cdot\vec{x}} M_{\kappa_2,i|\mu|}(z_2)h_{n}(\overline{y})\,\omega_2
%
  \end{split}
\end{equation}

\bigskip

Whereas the negative frequency modes (found via the charge conjugation of the above positive frequency modes, $V\equiv i \gamma^2 U^*$) are given by,
\begin{equation}
\begin{split}
V_{1,n}^{\rm in}=\frac{\gamma^0}{N_1a^{3/2}} \left\lbrace\left(i\partial_\eta-k_z\gamma^0 \gamma^3+aH\left(M \gamma^0 + L \gamma^0 \gamma^3 \right)\right)+\left(i\gamma^0 \gamma^2 \partial_y-\left(k_x-eBy\right)\gamma^0 \gamma^1\right)\right\rbrace \\  \times e^{iHLz}e^{-i\vec{k}\slashed{y}\cdot\vec{x}} W_{-\kappa_1,-i|\mu|}(-z_1)h_{n}(y_-) \,\omega_3
%
  \end{split}
\end{equation}
\begin{equation}
\begin{split}
V_{2,n}^{\rm in}=\frac{\gamma^0}{N_2a^{3/2}} \left\lbrace\left(i\partial_\eta-k_z\gamma^0 \gamma^3+aH \left(M \gamma^0 + L \gamma^0 \gamma^3 \right)\right)+\left(i\gamma^0 \gamma^2 \partial_y-\left(k_x-eBy\right)\gamma^0 \gamma^1\right)\right\rbrace \\  \times  e^{iHLz}e^{-i\vec{k}\slashed{y}\cdot\vec{x}} W_{-\kappa_2,-i|\mu|}(-z_2)h_{n}(y_-)\omega_4
%
   \end{split}
\end{equation}
\begin{equation}
\label{negative1out}
\begin{split}
V_{1,n}^{\rm out}=\frac{\gamma^0}{M_1a^{3/2}} \left\lbrace\left(i\partial_\eta-k_z\gamma^0 \gamma^3+aH\left(M \gamma^0 + L \gamma^0 \gamma^3 \right)\right)+\left(i\gamma^0 \gamma^2 \partial_y-\left(k_x-eBy\right)\gamma^0 \gamma^1\right)\right\rbrace \\  \times  e^{iHLz}e^{-i\vec{k}\slashed{y}\cdot\vec{x}} M_{-\kappa_1,-i|\mu|}(-z_1)h_{n}(y_-)\,\omega_3
%
 \end{split}
\end{equation}
\begin{equation}
\begin{split}
V_{2,n}^{\rm out}=\frac{\gamma^0}{M_2a^{3/2}} \left\lbrace\left(i\partial_\eta-k_z\gamma^0 \gamma^3+aH\left(M \gamma^0 + L \gamma^0 \gamma^3 \right)\right)+\left(i\gamma^0 \gamma^2 \partial_y-\left(k_x-eBy\right)\gamma^0 \gamma^1\right)\right\rbrace \\  \times  e^{iHLz}e^{-i\vec{k}\slashed{y}\cdot\vec{x}} M_{-\kappa_2,-i|\mu|}(-z_2)h_{n}(y_-)\,\omega_4
%
 \end{split}
\end{equation}

The normalisation constants, $N_1,\,N_2,\, M_1,\, M_2$ are given by \ref{nc}. We shall explicitly evaluate $N_1$ below. The rest can be derived in a similar manner. Using \ref{eigenkets} into \ref{positive1in}, we find after some algebra
\begin{eqnarray}
\int d^3x a^3\,U_{1}^{\dagger}{^\text{in}}U_{1}^\text{in}{^\prime}&=&\frac{1}{N_1^2}\int d^3x\,e^{-i(\vec{k}_\slashed{y}-\vec{k}_\slashed{y}^\prime)\cdot\vec{x}}\times\nonumber\\
&&\Bigg[\left(-i\partial_\eta{W_{\kappa_1,i|\mu|}(z_1)}^*-\left(\frac{k_zL}{\sqrt{M^2+L^2}}-aH\sqrt{M^2+L^2}\right)W_{\kappa_1,i|\mu|}^*(z_1)\right)h_{n}(\overline{y})\omega_1^{\dagger}\nonumber\\
&+&\frac{k_zM}{\sqrt{M^2+L^2}}X_1^{\dagger} W_{\kappa_1,i|\mu|}^*(z_1)h_{n}(\overline{y})-\left(\partial_y{h_{n}(\overline{y})}+\left(k_x+eBy\right) h_{n}(\overline{y})\right)W_{\kappa_1,i|\mu|}^*(z_1)\omega_3^{\dagger}\Bigg]\times\nonumber\\
&&\Bigg[\left(-i\partial_\eta{W_{\kappa_1,i|\mu|}(z_1)}-\left(\frac{k_zL}{\sqrt{M^2+L^2}}-aH\sqrt{M^2+L^2}\right)W_{\kappa_1,i|\mu|}(z_1)\right)h_{n}(\overline{y})\omega_1\nonumber\\
&+&\frac{k_zM}{\sqrt{M^2+L^2}}X_1 W_{\kappa_1,i|\mu|}(z_1)h_{n}(\overline{y})-\left(\partial_y{h_{n}(\overline{y})}+\left(k_x+eBy\right) h_{n}(\overline{y})\right)W_{\kappa_1,i|\mu|}(z_1)\omega_3\Bigg]
\label{N1}
\end{eqnarray}

The $x$ and $z$ integrals trivially give, $\delta(k_x-k'_x)\delta (k_z-k'_z)\equiv \delta^2(\vec{k}_\slashed{y}-\vec{k}_\slashed{y}^\prime)$. Using the orthonormality of $\omega_1$ and $\omega_3$, and the definition of the variable $\overline{y}$ appearing below \ref{dirac9'}, the $y$ integral is extracted to be, 
\[\int d \overline{y} \left[\partial_{y}{h_{n}(\overline{y})}\partial_{y} h_{n^\prime}(\overline{y})+\overline{y} h_{n}(\overline{y})\partial_{y} h_{n^\prime}(\overline{y})+\overline{y} h_{n^\prime}(\overline{y})\partial_{y} h_{n}(\overline{y})+\overline{y}^2h_{n^\prime}(\overline{y}) h_{n}(\overline{y})\right]\]\\
Using some properties of the Hermite polynomials~\cite{AS}, the first term equals
%
$$\int d\overline{y}\, \partial_{y}{h_{n}(\overline{y})}\partial_{y} h_{n^\prime}(\overline{y})=3eB\left(n+\frac16\right)\delta_{n n^\prime}$$
%
Second and third integrals vanish,
%
$$\int d\overline{y}\, \overline{y} h_{n}(\overline{y})\partial_{y} h_{n^\prime}(\overline{y}) =0=\int d\overline{y}\, \overline{y} h_{n'}(\overline{y})\partial_{y} h_{n}(\overline{y})$$
%
whereas the fourth integral equals,
%
$$\int d\overline{y}\, \partial_{y}{h_{n}(\overline{y})}\partial_{y} h_{n^\prime}(\overline{y}) y^2_ {+}   = eB \left(n+\frac12\right) \delta_{nn^\prime}$$
%
Collecting all the pieces, the $\overline{y}$ integral becomes
\begin{eqnarray}
\label{inthermite}
 4eB\left(n+\frac14\right) \delta_{nn^\prime}
\end{eqnarray}

Since normalisation is time independent, we may choose the arguments of the Whittaker functions in \ref{N1} as per our convenience.  Accordingly, we choose $\eta \to -\infty $, for which $W_{\kappa_1,i|\mu|}\approx e^{-z_1/2}z_{1}^{\kappa_1}$. We have 
\begin{eqnarray}
\label{Whit1}
&&W_{\kappa_1,i|\mu|}(z_1)(W_{\kappa_1,i|\mu|}(z_1))^*=
e^{\pi|\kappa_1| {\rm sgn}(k_z)}, \qquad \partial_\eta W_{\kappa_1,i|\mu|}(z_1)\partial_\eta (W_{\kappa_1,i|\mu|}(z_1))^*=e^{\pi|\kappa_1| {\rm sgn}(k_z)}(k_z^2+S_1)\nonumber\\
&&\partial_\eta (W_{\kappa_1,i|\mu|}(z_1))^* W_{\kappa_1,i|\mu|}(z_1)-\partial_\eta W_{\kappa_1,i|\mu|}(z_1) (W_{\kappa_1,i|\mu|}(z_1))^*=2i e^{\pi|\kappa_1| {\rm sgn}(k_z)}\sqrt{k_z^2+S_1}
\end{eqnarray}
Putting everything together in \ref{N1}, we find the normalisation integral becomes $\delta^2(\vec{k}_\slashed{y}-\vec{k}_\slashed{y}^\prime)\delta_{nn'}$, with the choice
\begin{eqnarray}
\label{normalization}
N_1=e^{ \pi|\kappa_1|{\rm sgn}(k_z)/2}
\end{eqnarray}

The normalisation for the other in modes can be found in a similar manner. 

For the normalisation of the out modes, we choose the integration hypersurface to be in the asymptotic future, $\eta \to 0^-$, for our convenience and use in this limit, 
$$M_{\kappa_1,i|\mu|}\approx\left(2i\eta\sqrt{k^2_z+S_1}\right)^{1/2+i|\mu|}$$
The rest of the calculations remains the same.

%

\bibliographystyle{cas-model2-names}

\bibliography{cas-refs}

\begin{thebibliography}{99}


\bibitem{bell_4}
A.~Einstein, B.~Podolsky and N.~Rosen, {\it Can Quantum-Mechanical Description of Physical Reality Be Considered Complete}, Phys. Rev. \textbf{777} (1935) 

\bibitem{Bell}
S.~Bell, {\it On the Einstein-Podolsky-Rosen paradox}, Physics\textbf{1}  195 (1964)

\bibitem{CHSH}
J.~F.~Clauser, M.~A.~Horne, A.~Shimony and R.~A.~Holt, {\it Proposed experiment to test local hidden-variable theories}, Phys.~Rev.~Lett.\textbf{23} 880 (1969) 

\bibitem{Werner:1989zz} 
  R.~F.~Werner,
  {\it Quantum states with Einstein-Podolsky-Rosen correlations admitting a hidden-variable model},
   Phys.\ Rev.\ A {\bf 40}, 4277 (1989)
   
  


\bibitem{Vidal:1998re} 
  G.~Vidal,
  {\it Entanglement monotones},
  J.\ Mod.\ Opt.\  {\bf 47}, 355 (2000)
  [arXiv:quant-ph/9807077]





 \bibitem{bell_3}
R.~Horodecki, P.~Horodecki and M.~Horodecki, {\it Violating Bell inequality by mixed spin-$\frac{1}{2}$ states:
necessary and sufficient condition}, Phys.~Lett.~A\textbf{200}  340 (1995)

\bibitem{seprability}
M.~Horodecki, P.~Horodecki and R.~Horodecki, {\it Separability of mixed states: Necessary and suffcient conditions}, Phys. Lett.A \textbf{223}, (1996) [arXiv:quant-ph/9605038]



\bibitem{bell1:2003}
 S.~ Yu, Z.~ Chen, J.~ Pan, Y.~D.~ Zhang, 
{\it Classifying N-qubit Entanglement via Bell's Inequalities
}, PRL\textbf{90}, 080401 (2003)
[arXiv:quant-ph/0211063]



\bibitem{bell_1}
P.~Y.~Chang, S.~K.~Chu and C.~P.~Ma,
{\it Bell's Inequality and Entanglement in Qubits},
JHEP 09  \textbf{100} (2017).  [1705.06444 [quant-ph]]


\bibitem{Aspect1}
A.~Aspect, P.~Grangier and G.~Roger, {\it Experimental Tests of Realistic Local Theories via Bell's Theorem},
Phys.~Rev.~Lett.{\bf47}, 460 (1981)

\bibitem{Aspect2}
A.~Aspect, J.~Dalibard and G.~Roger, {\it Experimental test of Bell's inequalities using time varying analyzers},
Phys. Rev. Lett.49, 1804 (1982)



\bibitem{NHB:1998zz} 
 N.~Gisin,  H.~B.~Pasquinucci, 
  {\it Bell inequality, Bell states and maximally entangled states for n qubits
}, Phys.\ Rev.\ A {\bf 246},  (1998)
[arXiv:quant-ph/9804045]

\bibitem{Mermin}
N.~D.~Mermin, {\it Extreme Quantum Entanglement in a Superposition of Macroscopically Distinct States}, Phys.~Rev.~Lett.{\bf 65}, 1838 (1990)

\bibitem{Klyshko}
A.~V.~Belinski and D.~N.~Klyshko, {\it Interference of light and Bell's theorem}, Physics-Uspekhi{\bf 36}, 653 (1993)


\bibitem{Plenio:2007zz} 
  M.~B.~Plenio and S.~Virmani,
  {\it An Introduction to entanglement measures},
  Quant.\ Inf.\ Comput.\  {\bf 7}, 1 (2007)
  [quant-ph/0504163]


\bibitem{NielsenChuang} 
	M.~A.~Nielsen and I.~L.~Chuang (2010), {\it Quantum Computation and Information Theory} (Cambridge university press, UK)



\bibitem{Monogamy}
H.~S.~Dhar, A.~K.~Pal, D.~Rakshit, A.~S.~De and U.~Sen, {\it Monogamy of quantum correlations - a review}, (2016)  [arXiv:1610.01069v1[quant-ph]]


\bibitem{BKAY:2017}
 B.~ Richter, K.~ Lorek, A.~ Dragan, Y.~ Omar , 
{\it Effect of acceleration on localized fermionic Gaussian states: from vacuum entanglement to maximally entangled states},	 Phys.Rev.D \textbf{95}, 076004 (2017)
[arXiv:quant-ph/0211063]

\bibitem{FuentesSchuller:2004xp} 
  I.~Fuentes-Schuller and R.~B.~Mann,
  {\it Alice falls into a black hole: Entanglement in non-inertial frames},
  Phys.\ Rev.\ Lett.\  {\bf 95}, 120404 (2005)
  [arXiv:quant-ph/0410172]



\bibitem{degradation:2015}
P.~M.~Alsing, I.~F.~Schuller, R.~B.~Mann, T.~E.~Tessier, 
{\it Entanglement of Dirac fields in non-inertial frames
}, Phys.~Rev.~A \textbf{74}, 032326  (2006)
[arXiv:quant-ph/0603269]


\bibitem{nper:2011} 
 N.~ Friis, P.~ Köhler, E. Martin-Martinez, R.~ A.~ Bertlmann,
  {\it Residual entanglement of accelerated fermions is not nonlocal}, Phys.Rev.A \textbf{84}  062111 (2011)
[arXiv:1107.3235[quant-ph]]

\bibitem{degradation:2015n}
 B.~ Richter, Y.~Omar
{\it Degradation of entanglement between two accelerated parties: Bell states under the Unruh effect
}, Phys.Rev.A \textbf{92}, 022334 (2015)
[arXiv:1503.07526[quant-ph]]




\bibitem{EE}
J. Maldacena and G. L. Pimentel, {\it Entanglement entropy in de Sitter space}, JHEP \textbf{1302}, 038 (2013) [arXiv:1210.7244 [hep-th]]
 
   \bibitem{Maldacena:2015bha} 
  J.~Maldacena,
{\it A model with cosmological Bell inequalities},
  Fortsch.\ Phys.\  {\bf 64}, 10 (2016)
  [arXiv:1508.01082 [hep-th]]

\bibitem{bell:2017}
S.~ Kanno, J.~Soda 
{\it Infinite violation of Bell inequalities in inflation
}, Phys.Rev.D \textbf{96}, 0211063  (2017)
[arXiv: 1705.06199 [hep-th]]

\bibitem{Fuentes:2010dt}
I.~Fuentes, R.~B.~Mann, E.~Martin-Martinez and S.~Moradi,
{\it Entanglement of Dirac fields in an expanding spacetime}
Phys. Rev. D \textbf{82}, 045030 (2010)
[arXiv:1007.1569[quant-ph]]

\bibitem{vaccum EE for fermions}
S.~Kanno, M.~Sasaki and T.~Tanaka, {\it Vacuum State of the Dirac Field in de Sitter Space and Entanglement Entropy}, JHEP \textbf{1703}, 068 (2017)  [arXiv:1612.08954 [hep-th]]

\bibitem{QC in deSitter}
J. Soda, S. Kanno and J. P. Shock,{\it Quantum Correlations in de Sitter Space}, Universe \textbf{3}, no. 1, 2 (2017)

\bibitem{SSS}
 S.~Bhattacharya, S.~Chakrabortty and S.~Goyal,
 {\it Emergent $\alpha$-like fermionic vacuum structure and entanglement in the hyperbolic de Sitter spacetime},  Eur.Phys.J.C 79  9,\textbf{799} (2019). [arXiv: 1812.07317[hep-th]] 
 
 \bibitem{Bhattacharya:2019zno}
S.~Bhattacharya, S.~Chakrabortty and S.~Goyal,
{\it Dirac fermion, cosmological event horizons and quantum entanglement},
Phys. Rev. D \textbf{101}, no.8, 085016 (2020)
[arXiv:1912.12272 [hep-th]]
    
  \bibitem{SHN:2020}
    S.~Bhattacharya,
    H.~Gaur and N.~Joshi,
    {\it Some measures for fermionic entanglement in the cosmological de Sitter spacetime
},  Phys.Rev.D \textbf{102}, 045017 (2020) [arXiv: 2006.14212[hep-th]]

\bibitem{Choudhury:2016cso} 
  S.~Choudhury, S.~Panda and R.~Singh,
{\it Bell violation in the Sky},
  Eur.\ Phys.\ J.\ C {\bf 77}, no. 2, 60 (2017)
  [arXiv:1607.00237[hep-th]]

\bibitem{Choudhury:2017bou} 
  S.~Choudhury and S.~Panda,
{\it Entangled de Sitter from stringy axionic Bell pair I: an analysis using Bunch-Davies vacuum},
  Eur.\ Phys.\ J.\ C {\bf 78}, no. 1, 52 (2018)
  [arXiv:1708.02265[hep-th]]






  

  

   
   



 
\bibitem{Parker:2009uva}
L.~E.~Parker and D.~J.~Toms,
{\it Quantum Field Theory in Curved Spacetime: Quantized Field and Gravity},
Cambridge Univ. Press (2009)

\bibitem{Ebadi:2014ufa} 
  Z.~Ebadi and B.~Mirza,
  {\it Entanglement Generation by Electric Field Background},
  Annals Phys.\  {\bf 351}, 363 (2014)
  [arXiv:1410.3130[quant-ph]]

\bibitem{Li:2016zyv}
Y.~Li, Y.~Dai and Y.~Shi,
{\it Pairwise mode entanglement in Schwinger production of particle-antiparticle pairs in an electric field},
Phys. Rev. D \textbf{95}, no.3, 036006 (2017)
[arXiv:1612.01716[hep-th]]

\bibitem{Agarwal:2016cir}
A.~Agarwal, D.~Karabali and V.~Nair,
{\it Gauge-invariant Variables and Entanglement Entropy}
Phys. Rev. D \textbf{96}, no.12, 125008 (2017)
[arXiv:1701.00014[hep-th]]
 
\bibitem{Li:2018twv}
Y.~Li, Q.~Mao and Y.~Shi,
{\it Schwinger effect of a relativistic boson entangled with a qubit},
Phys. Rev. A \textbf{99}, no.3, 032340 (2019)
[arXiv:1812.08534[hep-th]]

\bibitem{Karabali:2019ucc}
D.~Karabali, S.~Kurkcuoglu and V.~Nair,
{\it Magnetic Field and Curvature Effects on Pair Production II: Vectors and Implications for Chromodynamics}
Phys. Rev. D \textbf{100}, no.6, 065006 (2019)
[arXiv:1905.12391[hep-th]]

\bibitem{Dai:2019nzv}
D.~C.~Dai,
{\it State of a particle pair produced by the Schwinger effect is not necessarily a maximally entangled Bell state},
Phys. Rev. D \textbf{100}, no.4, 045015 (2019)
[arXiv:1908.01005[hep-th]]

\bibitem{HSSS:2020}  
S.~Bhattacharya, S.~Chakrabortty, H.~Hoshino and S.~Kaushal,
{\it Background magnetic field and quantum correlations in the Schwinger effect}, 
Phys. Lett. B \textbf{811}, 135875 (2020)[arXiv:2005.12866[hep-th]]


 \bibitem{Balasubramanian}
 V.~ Balasubramanian,  M.~ B.~ McDermott and  M.~V.~ Raamsdonk,
 {\it Momentum-space entanglement and renormalization in quantum field theory
}, Phys.Rev.D \textbf{86}  045014 (2012) [arXiv:1108.3568[hep-th]]


 \bibitem{Momentum space entanglement} 
 G.~Grignani and G.~ W.~ Semenoff, {\it Scattering and momentum space entanglement}, Phys. Lett. B \textbf{772},   699-702 (2017)	[arXiv:1612.08858[hep-th]]


\bibitem{Ryu:2006bv} 
  S.~Ryu and T.~Takayanagi,
  {\it Holographic derivation of entanglement entropy from AdS/CFT},
  Phys.\ Rev.\ Lett.\  {\bf 96}, 181602 (2006)
  [hep-th/0603001]



\bibitem{Ryu:2006ef} 
  S.~Ryu and T.~Takayanagi,
  {\it Aspects of Holographic Entanglement Entropy},
  JHEP {\bf 0608}, 045 (2006)
  [hep-th/0605073]
  
  
 

  
  \bibitem{Boyanovsky:2018soy}
D.~Boyanovsky,
{\it Imprint of entanglement entropy in the power spectrum of inflationary fluctuations},
Phys. Rev. D \textbf{98}, no.2, 023515 (2018)
[arXiv:1804.07967[astro-ph.CO]]

  \bibitem{Rauch:2018rvx}
D.~Rauch, J.~Handsteiner, A.~Hochrainer, J.~Gallicchio, A.~S.~Friedman, C.~Leung, B.~Liu, L.~Bulla, S.~Ecker and F.~Steinlechner, \textit{et al.}
{\it Cosmic Bell Test Using Random Measurement Settings from High-Redshift Quasars},
Phys. Rev. Lett. \textbf{121}, no.8, 080403 (2018)
[arXiv:1808.05966[quant-ph]].

\bibitem{Morse:2020mdc}
M.~J.~P.~Morse,
{\it Statistical Bounds on CMB Bell Violation},
[arXiv:2003.13562 [astro-ph.CO]]


\bibitem{Subramanian:2015lua}
K.~Subramanian,
{\it The origin, evolution and signatures of primordial magnetic fields},
Rept.~Prog.~Phys.\textbf{79}, no.7, 076901 (2016)
[arXiv:1504.02311[astro-ph.CO]]

\bibitem{vilenkin}
M.~B.~Fr\"{o}b, J.~Garriga, S.~Kanno, M.~Sasaki, J.~Soda, T.~Tanaka and A.~Vilenkin, {\it Schwinger effect in de Sitter space}, JCAP 04 (2014) \textbf{009}  [arXiv:1401.4137[hep-th]]

\bibitem{Xue:2017}
T.~ Kobayashi, N.~ Afshordi,
{\it Schwinger Effect in 4D de Sitter Space and Constraints on Magnetogenesis in the Early Universe
},
JHEP \textbf{10}  166 (2014)
[arXiv:1408.4141[hep-th]]


\bibitem{Xue:2017cex}
C. Stahl, E. Strobel, S.~S~.Xue,
{\it Fermionic current and Schwinger effect in de Sitter spacetime},
Phys.Rev.D \textbf{93}, 025004 (2016)
[arXiv:1507.01686[gr-qc]]




\bibitem{Xue:2017ecx}
E.~ Bavarsad, C.~Stahl, S.~S.~ Xue,
{\it Scalar current of created pairs by Schwinger mechanism in de Sitter spacetime},
Phys. Rev. D \textbf{94}, (2016)
[arXiv:1602.06556[hep-th]]

\bibitem{yoko:2016tty}
T.~ Hayashinaka, T.~ Fujita, J.~ Yokoyama,
{\it Fermionic Schwinger effect and induced current in de Sitter space},
JCAP \textbf{07},101, 04165v2 (2016)
[arXiv:1603.04165[hep-th]]

        
\bibitem{Bavarsad:2017oyv} 
  E.~Bavarsad, S.~P.~Kim, C.~Stahl and S.~S.~Xue,
  {\it Effect of a magnetic field on Schwinger mechanism in de Sitter spacetime},
  Phys.\ Rev.\ D {\bf 97},
   025017 (2018)
  [arXiv:1707.03975[hep-th]]


\bibitem{Mironov:2011hp}
A.~Mironov, A.~Morozov and T.~N.~Tomaras,
{\it Geodesic deviation and particle creation in curved spacetimes},
Pisma~Zh.~Eksp.~Teor.~Fiz.\textbf{94}, 872 (2011)
[arXiv:1108.2821 [gr-qc]]





\bibitem{AS}
	M.~Abramowitz and I.~Stegun,
{\it Handbook of Mathematical Functions with Formulas, Graphs, and Mathematical Tables}, National Bureau of Standards (USA) (1964)







\bibitem{Allen}
B.~Allen, {\it Vacuum States in de Sitter Space}, Phys.~Rev.~D{\bf32}, 3136 (1985)

\bibitem{alpha-vaccuum1}
 J.~de~Boer, V.~Jejjala and D.~Minic, {\it Alpha-states in de Sitter space}, Phys. Rev. D \textbf{71}, 044013 (2005) [hep-th/0406217]
 
 \bibitem{Mottola:1984ar}
E.~Mottola,
{\it Particle Creation in de Sitter Space},
Phys.~Rev.~D\textbf{31}, 754 (1985)



\bibitem{Einhorn}
M.~B.~Einhorn and F.~Larsen, {\it Interacting quantum field theory in de Sitter vacua}, Phys.~Rev.~D{\bf67}, 024001 (2003) [hep-th/0209159]


\bibitem{Bedroya}
 A.~Bedroya, {\it de Sitter Complementarity, TCC, and the Swampland}, arXiv:2010.09760 [hep-th]




\bibitem{alpha-vacua}
H.~Collins, {\it Fermionic alpha-vacua}, Phys. Rev. D \textbf{71}, 024002 (2005) [hep-th/0410229]

 
 \bibitem{EE for alpha1}
N. Iizuka, T. Noumi and N. Ogawa, {\it Entanglement entropy of de Sitter space $\alpha$-vacua}, Nucl. Phys. B \textbf{910}, 23 (2016) [arXiv:1404.7487[hep-th]]

\bibitem{EE for alpha}
S. Kanno, J. Murugan, J. P. Shock and J. Soda, {\it Entanglement entropy of $\alpha$-vacua in de Sitter space},
JHEP \textbf{1407}, 072 (2014) [arXiv:1404.6815[hep-th]]

\bibitem{BI for mixed state}
Y.~Kwon, {\it No survival of Nonlocalilty of fermionic quantum states with
alpha vacuum in the infinite acceleration limit}, Phys. Lett. B \textbf{748} (2015)

\bibitem{Choudhury:2017qyl} 
  S.~Choudhury and S.~Panda,
{\it Quantum entanglement in de Sitter space from stringy axion: An analysis using $\alpha$ vacua},
  Nucl.\ Phys.\ B {\bf 943}, 114606 (2019)
  [arXiv:1712.08299[hep-th]]



\bibitem{Gabriel:1999yz} 
  C.~Gabriel and P.~Spindel,
  {\it Quantum charged fields in Rindler space},
  Annals Phys.\  {\bf 284}, 263 (2000)
  [gr-qc/9912016]




\end{thebibliography}


\end{document}